\documentclass[aps,prb,twocolumn,superscriptaddress,floatfix]{revtex4}
\usepackage{graphicx,epsf}
\usepackage{amsmath}
\usepackage{verbatim}
\usepackage{float}
\pdfoutput=1

\begin{document}

\title{
Finite-temperature spectra and quasiparticle interference in Kondo lattices:\\
From light electrons to coherent heavy quasiparticles
}

\author{Adel Benlagra}
\affiliation{Institut f\"ur Theoretische Physik, Technische Universit\"at Dresden,
01062 Dresden, Germany}
\author{Thomas Pruschke}
\affiliation{Institut f\"ur Theoretische Physik, Georg-August-Universit\"at G\"ottingen,
Friedrich-Hund-Platz 1, 37077 G\"ottingen, Germany}
\author{Matthias Vojta}
\affiliation{Institut f\"ur Theoretische Physik, Technische Universit\"at Dresden,
01062 Dresden, Germany}

\date{September 1, 2011}

\begin{abstract}
Recent advances in scanning tunneling spectroscopy performed on heavy-fermion metals
provide a window onto local electronic properties of composite heavy-electron
quasiparticles. Here we theoretically investigate the energy and temperature evolution of
single-particle spectra and their quasiparticle interference caused by point-like
impurities in the framework of a periodic Anderson model. By numerically solving
dynamical-mean-field-theory equations, we are able to access all temperatures and to
capture the crossover from weakly interacting $c$ and $f$ electrons to fully coherent
heavy quasiparticles. Remarkably, this crossover occurs in a dynamical fashion at an
{\em energy-dependent} crossover temperature. We study in detail the associated Fermi-surface
reconstruction and characterize the incoherent regime near the Kondo temperature.
Finally, we link our results to current heavy-fermion experiments.
\end{abstract}

\maketitle


\newcommand{\urusi} {URu$_2$Si$_2$}

\newcommand{\hybb}  {g_0}               
\newcommand{\hyb}   {g}                 
\newcommand{\epsb}  {\varepsilon_0}     
\newcommand{\eps}   {\varepsilon}       
\newcommand{\Jb}    {J_{\rm K}}         
\newcommand{\vvb}   {u_0}               
\newcommand{\vv}    {u}                 
\newcommand{\TK}    {T_{\rm K}}
\newcommand{\Tcoh}  {T_{\rm coh}}
\newcommand{\JK}    {J_{\rm K}}
\newcommand{\Jc}    {J_{c}}

\newcommand{\im}    {{\rm Im}}
\newcommand{\re}    {{\rm Re}}

\newcommand{\bea}{\begin{eqnarray}}
\newcommand{\eea}{\end{eqnarray}}
\newcommand{\beq}{\begin{equation}}
\newcommand{\eeq}{\end{equation}}
\newcommand{\benu}{\begin{enumerate}}
\newcommand{\enu}{\end{enumerate}}
\newcommand{\la}{\langle}
\newcommand{\ra}{\rangle}
\newcommand{\al}{\alpha}
\newcommand{\be}{\beta}
\newcommand{\om}{\omega}
\newcommand{\Om}{\Omega}
\newcommand{\ep}{\epsilon}
\newcommand{\ta}{\tau}
\newcommand{\si}{\sigma}
\newcommand{\dl}{\delta}
\newcommand{\lam}{\lambda}
\newcommand{\ham}{\mathcal{H}}
\newcommand{\ord}{\mathcal{O}}
\newcommand{\prm}{\prime}
\newcommand{\ptl}{\partial}
\newcommand{\sgn}{{\rm Sgn}}
\newcommand{\tg}{\gamma}
\newcommand{\tG}{\tilde{G}}
\newcommand{\G}{\mathcal{G}}
\newcommand{\tOm}{\tilde{\Omega}}
\newcommand{\cda}{c^{\dagger}}
\newcommand{\fda}{f^\dagger}
\newcommand{\bk}{{\bf k}}
\newcommand{\bq}{{\bf q}}
\newcommand{\br}{{\bf r}}
\newcommand{\fullint}{\int_{- \infty}^{\infty}}
\newcommand{\et}{\eta_{\Omega}}
\newcommand{\fo}{F_0^{\sigma}}
\newcommand{\tst}{T^{\ast}}
\newcommand{\ek}{\epsilon_{\bk}}
\newcommand{\nn}{\nonumber \\}


\section{Introduction}

Heavy-fermion metals,\cite{hewson,colemanrev} where strongly localized $f$ electrons tend to form magnetic
moments which interact with more delocalized conduction ($c$) electrons, constitute an
active and fascinating research area in condensed matter physics: Many non-trivial
phenomena like competing orders, quantum criticality, unconventional superconductivity,
and quantum Griffiths phases find their realization in members of the heavy-fermion
family.\cite{stewart01,hvl}
Despite significant theoretical progress in the field of correlated electrons,
the rich physics of heavy-fermion materials remains only partially understood. The
difficulties lie with spatially non-local phenomena -- like exotic magnetism and
superconductivity -- while we have at least a qualitative understanding of the local
process of temperature-dependent heavy-quasiparticle formation.

Experimentally, most investigations have concentrated on thermodynamic or transport
properties, which are accessible in a straightforward fashion. In
contrast, experimental results using single-electron spectroscopy in the heavy-fermion regime
are scarce, as the required energy resolution in the sub-meV range is difficult to reach
with present-day photoemission techniques.

Spectroscopic-imaging scanning tunneling microscopy (SI-STM) is a surface-sensitive probe
which allows to measure single-particle spectra with the required energy resolution.
Recent efforts in preparing clean surfaces of layered heavy-fermion compounds have
allowed for the first time to measure spatially resolved maps of the differential
conductance, $dI/dV$, in heavy-fermion compounds.\cite{lee_expt,davis_urusi,
yazdani_urusi,wirth} Thanks to impurity scattering processes, such imaging also allows to
partially reconstruct momentum-space information on the electronic spectra, via
energy-dependent Friedel oscillations, dubbed ``quasiparticle interference''
(QPI).\cite{crommie93,hoffman02,mcelroy03,qpi_cup,wanglee03} Thus, SI-STM provides a unique
opportunity to visualize the many-body quantum physics generated by the coupling between
$f$ and $c$ electrons.
The experimental data obtained on \urusi\cite{davis_urusi,yazdani_urusi} show a periodic
lattice of Fano-shaped tunneling spectra at lowest temperatures, which is likely to arise
from a combination of the hybridization between $c$ and $f$ bands and interference
effects of different tunneling paths. In addition, the QPI signal shows signatures of
heavy-band formation near the Fermi level.

The theoretical modeling of these tunneling spectra has so far been restricted to a few
easily accessible limits. Slave-boson mean-field techniques have been applied to the
Kondo lattice model,\cite{coleman09,morr10} which describe elements of the low-energy and
low-temperature physics of heavy quasiparticles: The mean-field Hamiltonian is that of
two hybridized bands of otherwise non-interacting fermions. As all inelastic scattering
processes are neglected, physical properties at elevated energies or temperatures cannot
be described: Among the artifacts are a hard hybridization gap in the heavy-fermion state
and an artificial finite-temperature transition between this and a decoupled
high-temperature state. A model of heavy quasiparticle bands has been supplied by
phenomenological Fermi-liquid self-energies in Ref.~\onlinecite{woelfle10} in order to
capture the partial filling of the hybridization gap. Such a quasiparticle broadening has
been shown to be essential to the observation of the Fano-shaped peak in a Kondo lattice
system. However, a consistent modeling for all energies and temperatures is not yet available.

The purpose of this paper is to close this gap: We provide a detailed study of
temperature-dependent electronic spectra and QPI phenomena in a generic model of
heavy-fermion metals, the periodic Anderson model (PAM). To this end, we numerically
solve dynamical mean-field theory (DMFT) equations using Wilson's numerical
renormalization group (NRG), a method which provides real-frequency spectra at all
temperatures. The calculations allow us to track the formation of coherent heavy
quasiparticles as function of energy and temperature, thus providing a basis for the
interpretation of SI-STM experiments. As detailed below, we find that the crossover
temperature associated with the Kondo-driven Fermi-surface reconstruction is
energy-dependent, demonstrating the dynamical character of the screening process.

\subsection{Outline}

The paper is organized as follows. In Sec.~\ref{sec:model} we introduce the PAM and its
treatment within DMFT using the NRG impurity solver. Sec.~\ref{sec:tunn} describes the
calculational scheme for the tunneling and QPI signals, the latter involving the
scattering off isolated impurities. In Sec.~\ref{sec:scales} we discuss various general
aspects of heavy fermions and their description within a local self-energy approximation,
with focus on energy scales and corresponding features in the single-particle spectra.
Sec.~\ref{sec:res} is devoted to a detailed discussion of our numerical results, where we
present both momentum-integrated and momentum-resolved single-particle spectra as well as
constant-energy cuts through single-particle and QPI spectra, all as function of
temperature. This will allow for a detailed discussion of the crossover from weakly
interacting $c$ and $f$ bands at high $T$, with a small Fermi surface, to coherent heavy
quasiparticles with a large Fermi surface at low $T$. A summary of experimentally
relevant aspects closes the paper.

Readers mainly interested in the temperature-driven band reconstruction from light to
heavy quasiparticles should jump ahead to Sec.~\ref{subsec:bands} and
Fig.~\ref{results:renormalized_bands}.


\section{Model and dynamical mean-field approximation}
\label{sec:model}

Here we describe the microscopic approach taken to calculate single-particle spectra. The
concrete modeling of SI-STM data is covered in Sec.~\ref{sec:tunn}.

\subsection{Periodic Anderson model}
\label{subsec:PAM}

A standard microscopic model used in the context of heavy-fermion compounds is
the so-called periodic Anderson (or Anderson lattice) model,\cite{hewson} which describes hybridized
$c$ and $f$ bands with a strong local repulsion on the $f$ orbitals:
\bea
\ham = && \sum_{\bk \si} \left [ \ek \cda_{\bk \si} c_{\bk \si} + \ep_f \fda_{\bk \si} f_{\bk \si} +
V_\bk ( \cda_{\bk \si} f_{\bk \si} + \mbox{H.c.} )  \right ]\nn
&& +  U \sum_i n^f_{i\downarrow} n^f_{i \uparrow}.
\label{PAM}
\eea
Here $\cda_{\bk \si} (\fda_{\bk \si})$ create a conduction electron ($f$-electron) of
momentum $\bk$ and spin $\si$, and $n^f_{i\si} = \fda_{i\si} f_{i\si}$ counts the $f$
particles at site $i$. $\ek$ and $\ep_f$ are the band energies of the conduction and
dispersionless $f$-electrons, respectively, measured relative to the chemical potential. Finally,
$V_\bk$ is the hybridization between the two fermion bands, and $U$ is the local Coulomb
repulsion between the $f$-electrons, which is often to be taken the largest energy scale.

For negative $\ep_f$ and large $U$, local moment tends to form on the $f$ orbitals. The
charge fluctuation scale of the $f$-electrons is determined by the Anderson width
$\Gamma_0= (1/\mathcal N) \sum_\bk V_\bk^2 \delta(\ek)$. Provided that $\Gamma_0 \ll U$,
a mixed-valence regime is reached for $|\ep_f| \sim \Gamma_0$ with $0 < n_f < 1$, whereas
$|\ep_f| \gg \Gamma_0$ leads to stable local moments. In this regime, charge fluctuations
can be integrated out, and one obtains the Kondo lattice model
\beq
\ham_{\rm KLM}=\sum_{\bk \si} \ek  \cda_{\bk \si} c_{\bk \si} + \JK \sum_i \vec S_i \cdot \vec s_i,
\label{KLM}
\eeq
where the impurity spin $\vec S_i$ is coupled to the conduction electrons spin density at
site i, $\vec s_i=\sum_{\si \si^\prime}\cda_{i \si} \vec \tau_{\si \si^\prime} c_{i
\si^\prime}/2$ and $c_{i \si}=(1/\mathcal{N})\sum_{\bk} e^{i \br_i \cdot \bk}c_{\bk \si}$.
For a local hybridization, $V_\bk\equiv V$, the Kondo coupling $\JK$ in \eqref{KLM} is
related to the parameters of the Anderson model through
\beq
\JK = 2 V^2 \left ( \frac{1}{| \ep_f|} + \frac{1}{|\ep_f + U|} \right ).
\eeq

The Anderson model \eqref{PAM} describes orbitally non-degenerate $f$ states -- this situation
often applies to Ce-based heavy-fermion systems: the lowest $f$ configuration is $f^1$,
whose multiplicity is reduced to a Kramers doublet due to crystalline-electric-field
(CEF) splitting, and which is well separated from the $f^0$ state. The situation is more
complicated, e.g., in uranium compounds where $f^2$ configurations cannot be neglected.
It is, however, believed that Eq.~\eqref{PAM} captures, at least qualitatively, the
low-energy physics of many heavy-fermion compounds.

\subsection{Dynamical mean-field theory}
\label{subsec:dmft}

As the periodic Anderson model with non-zero $U$ is not exactly solvable, further
approximations have to be made. A well-established and successful method, which is able
to capture the effects of strong local correlations, is the so-called dynamical mean-field
theory.\cite{dmft_metzner, dmft_georges} Here, the self-energy due to the
Hubbard-like interaction $U$ is assumed to be momentum-independent,
\begin{equation}
\Sigma_f(\bk, \om) \rightarrow \Sigma_f(\om).
\label{SE_property}
\end{equation}
This approximation becomes exact in the limit of infinite coordination number. As a
result of Eq.~\eqref{SE_property}, the lattice problem Eq.~\eqref{PAM} can be mapped onto
an effective single-impurity Anderson model supplemented by a self-consistency condition
for the effective medium (or bath) of the impurity. Assuming that both $c$ and $f$
electrons live on the same lattice structure, with a local hybridization $V_\bk\equiv V$,
the self-consistency equation for the single-particle propagator reads
\bea
\label{PAM_self_consistency}
\G^0_{ff}(i \om) &\equiv& \frac{1}{\mathcal N} \sum_{\bk} \frac{1}{i \om-\ep_f-\Sigma_f(i \om)- \frac{V^2}{i
\om-\ek}}\\
&=&\int d \ep \frac{\rho_0(\ep)}{i \om-\ep_f-\Sigma_f(i \om)- \frac{V^2}{i \om-\ep}} \nn
&=& \frac{1}{i\om-\ep_f-\tilde \Delta(i \om)-\Sigma_f(i \om)}= \G^{\rm SIAM}(i \om),
\nonumber
\eea
where $\G^0_{ff}(i \om)$ is the local $f$ Green's function, with the superscript $0$
denoting the impurity-free system (see below), and $\rho_0(\ep) = (1/\mathcal N) \sum_\bk
\delta(\ep-\ek)$ is the non-interacting $c$ electron density of states (DOS).
Further, $\tilde \Delta(i\om)$ is a generalized hybridization function, which define the
impurity problem and depends implicitly on the $f$ self-energy $\Sigma_f(i\om)$ and is
thus different from its non-interacting form due to the $f$ correlations effects.

Using the $f$ self-energy one can express the momentum-dependent Green's functions as
follows:
\bea
\label{greens_cc}
\G^0_{cc}(\bk, i \om)& = &\frac{i \om - \ep_f - \Sigma_f(i \om)}{  (i \om - \ek)(i \om - \ep_f - \Sigma_f(i \om)) -V^2},\\
\label{greens_ff}
\G^0_{ff}(\bk, i \om)& = &\frac{i \om - \ek}{  (i \om - \ek)(i \om - \ep_f - \Sigma_f(i \om)) -V^2},\\
\label{greens_cf}
\G^0_{cf}(\bk, i \om)& = &\frac{V}{  (i \om - \ek)(i \om - \ep_f - \Sigma_f(i \om)) -V^2}.
\label{greens_matrix}
\eea
Here, $\G^0_{cc}(\bk, i \om)$ is the Fourier transform of $\G^0_{cc}(\bk, \tau) = - \langle T_\tau
c_{\bk\sigma}(\tau) c_{\bk\sigma}^\dagger \rangle$, with $T_\tau$ being the time-ordering operator
on the imaginary time axis; $\G^0_{ff}(\bk, i \om)$ and $\G^0_{cf}(\bk, i \om)$ are
defined analogously. Note that all Green's functions are independent of spin in the
considered paramagnetic state.
From Eq. \eqref{greens_cc}, we can define a self-energy $\Sigma_c(i \om)$ for the
conduction electrons according to
\begin{equation}
\Sigma_c(i \om) \equiv \frac{V^2}{i \om - \ep_f - \Sigma_f(i \om)}.
\end{equation}


\subsection{Numerical renormalization group}

To solve the effective impurity problem arising within DMFT or its generalizations, a
variety of different methods have been developed, all with individual advantages and
drawbacks. In the present case, a non-perturbative method is preferable which (i) can
access the small energies and temperatures relevant for heavy-fermion formation and (ii)
directly provides spectral function on the real frequency axis, such that analytic
continuation from the imaginary axis is not required.

Here, we choose Wilson's numerical renormalization group (NRG) technique,\cite{NRGrev}
which has been successfully implemented in the context of DMFT for the Hubbard model, for
periodic Anderson model, and the Kondo lattice
model.\cite{IS_sakai,bulla99,bulla00,pruschke00,shimizu00} NRG is based on a logarithmic
discretization of the energy axis, controlled by a parameter $\Lambda > 1$: The energy
axis is divided into intervals $[\pm W \Lambda^{-(n+1)}, \pm W \Lambda^{-n}]$ for $n=0,1,
\ldots, \infty$, where $W$ is the half-bandwidth of the bare conduction band. The
original Hamiltonian is then mapped onto a semi-infinite chain, the Wilson chain, where
the $N$th link represents an exponentially decreasing energy scale $\sim \Lambda^{-N/2}$.
The chain Hamiltonian is diagonalized iteratively, by starting from the impurity and
successively adding chain sites. In each step, the high-energy states are truncated to
maintain a manageable number of states $N_s$.  The reduced states are expected to capture
the spectrum of the original Hamiltonian on a scale $\sim \Lambda^{-N/2}$, corresponding
to a temperature $T = W/\bar{\beta} \Lambda^{-N/2}$ from which all thermodynamic
expectation values are calculated. Spectral information at the temperature $T$ is
calculated by collecting information from NRG steps $1,\ldots, N$ which yields discrete
spectra on a logarithmic energy scale down to energies of (a fraction of) $T$.
The impurity self-energy needed in the DMFT iteration is most accurately evaluated as the
ratio of two propagators, eliminating the need for using the Dyson equation.\cite{bulla98}
For more details of NRG and its application to DMFT we refer the reader to a recent review
article.\cite{NRGrev}

The DMFT-NRG method has been employed to investigate heavy-fermion physics in the
past.\cite{pruschke00,shimizu00,costi02,grenzebach06,grenzebach08,bodensiek11} However, those investigations
mainly focused on thermodynamic properties or momentum-integrated spectra, and
consequently the Bethe lattice was used.
Here, we are motivated by layered heavy-fermion materials being good candidates for STM
investigations, and thus we perform all numerical calculations for a two-dimensional (2d)
square lattice with nearest-neighbor hopping. NRG parameters are
$\Lambda=2$, $N_s=600$, and $\bar{\beta}=1$ unless otherwise noted.



\section{Tunneling spectra and quasiparticle interference}
\label{sec:tunn}

After having described our theoretical approach to the clean (i.e. impurity-free) bulk
heavy-fermion system, we now turn to aspects relevant to SI-STM.

\subsection{Tunneling spectra}

In the context of tunneling experiments both on single magnetic impurities on metallic
surfaces\cite{madhavan98,kroha00}
and on heavy-fermion systems,\cite{coleman09,morr10} it has been proposed that the
tunneling current arises from the interference of two channels, namely a tunneling
process from the tip -- positioned over a particular site $\br_i$ -- into a
conduction-electron state, with an amplitude $t_c$, and another one to an $f$-electron
state, with an amplitude $t_f$.
Assuming both processes to be spatially local, the
tunneling piece of the Hamiltonian may be written as:
\begin{equation}
\ham_T=\sum_{\si} \left ( t_c \cda_{i \si} + t_f  \fda_{i \si}\right ) p_{\si} + \mbox{H.c.},
\label{tip_hamiltonian}
\end{equation}
where $p_{\si}$ destroys an electron with spin $\si$ in the tip.

Assuming further that the tip and the system are in thermal equilibrium, \cite{te_note} the total
tunneling current, flowing from the tip into the system, to lowest order in the tunneling
amplitudes $t_f, t_c$, is  given by
\begin{eqnarray}
I(\br_i, eV) = \frac{2e}{\hbar}\int d \om \,\, \rho_{\rm tip}(\om - e V) \left[ f(\om-eV) -f(\om)\right] \times \nn
\hspace{-0.3cm} {\im} \left [t_c^2 \G_{cc}(\br_i,\br_i, \om) + t_f^2 \G_{ff}(\br_i,\br_i, \om) + 2 t_c t_f  \G_{cf}(\br_i,\br_i, \om)\right ].\nn
\label{tunneling_current}
\end{eqnarray}
Here, $\rho_{\rm tip}$ is the tip DOS, $V$ the applied bias voltage,
$f(\om)=1/(1+e^{\om/T})$ the Fermi-Dirac function, and all
real-frequency Green's functions are to be understood as retarded ones.
In a translation-invariant system, the real-space Green's function $\G_{cc}(\br_i,\br_j,
\om)$ depends on $(\br_i-\br_j)$ only, such that $I(\br_i, eV)$ is independent of the tip
position $\br_i$. (Note that we do not account for the sub-atomic structure of the Bloch
states.)
The last term in \eqref{tunneling_current} describes the quantum interference processes
between the two tunneling paths.

If the tip DOS $\rho_{\rm tip}(\om)$ is independent of $\om$, the differential tunneling
conductance reduces to
\begin{eqnarray}
&&\frac{dI}{dV}(\br_i, eV) = - \rho_{\rm tip} \frac{2 e}{\hbar} \int  d \om \,\, f^\prime (\om-eV)\times \nn
&&~~~~~~~~~~{\im} \big [t_c^2 \G_{cc}(\br_i,\br_i, \om) + t_f^2 \G_{ff}(\br_i,\br_i, \om) \nn
&&~~~~~~~~~~~~~~~~~~~~~~~~~~~~~~+ 2 t_c t_f  \G_{cf}(\br_i,\br_i, \om)\big ],
  \label{tunneling_conductance}
\end{eqnarray}
which is the quantity of interest in discussing STM data.



\subsection{Impurities}
\label{subsec:impurities}

Friedel oscillations and QPI phenomena rely on the breaking of translational invariance
due to impurities or crystal imperfections which act as scattering centers. In fact,
intentional impurity doping has been employed to enhance QPI signatures in SI-STM
experiments.\cite{davis_urusi}

As we are interested in qualitative features, we use the simplest model capable of
describing interference phenomena in the framework of the Anderson model. Consider a
random distribution of $n_s$ non-magnetic point-like scatterers at sites ${ \bf R}_i,
i=1, \ldots, n_s$ which couple to the conduction electrons only. This introduces a
scattering term $\ham_S$ in the Hamiltonian
\bea
\mathcal \ham_s &=&  g_s \sum_{i = 1, \si}^{n_s} \cda_\si ({ \bf R}_i) c_\si ({ \bf R}_i)\nn
&=&\sum_{i = 1, \si}^{n_s} (\cda_{\si}({ \bf R}_i) \,\,\, \fda_{\si}({ \bf R}_i)) \,\,\hat g_s \left (
 \begin{array}{c}
c_{\si}({ \bf R}_i)\\
f_{\si}({ \bf R}_i)
\end{array}
\right )
\eea
where $g_s$ is the strength of the impurity potential and
\beq
\hat g_s = \left (
 \begin{array}{cc}
g_s & 0\\
0 & 0
\end{array}
\right ).
\label{gs}
\eeq
Apparently, more general forms of $\hat g_s$ are possible, but will not be considered
here.

For static point-like impurities embedded into a free-electron gas ($U=0$ in our case),
the Green's function matrix $\hat\G$ in the presence of scattering can be expressed
exactly via the T-matrix as follows:
\bea
\label{gf_t_matrix}
\hat \G(\br, \br^\prime, \om)&=& \hat \G^0(\br-\br^\prime, \om) \\
&+ &\sum_{i, j=1}^{n_s}\hat \G^0(\br-{\bf R}_i, \om)\hat  T_{ij}(\om) \hat \G^0({\bf R}_j-\br^\prime, \om)\nonumber
\eea
with
\beq
\hat T_{ij}= \hat t \,\,\delta_{ij} + \sum_{l=1}^{n_s} \hat t \left [1-\delta_{l, i} \right ]\hat \G^0({\bf R}_i-{\bf R}_l) \hat T_{lj}
\label{t_matrix}
\eeq
and the single-impurity T-matrix $\hat t(\om)$
\beq
\hat t(\om)= \hat g_s \left [ \hat 1 - \hat g_s  \hat \G^0(\br=0, \om) \right ]^{-1}
\label{1_t_matrix}
\eeq
where $\hat 1$ is the identity matrix in spatial, spin and fermion indices,
and $\hat \G^0$ now denotes Green's functions in the absence of scattering.

\begin{figure}[!t]
\begin{tabular}{ll}
a)&\\
&\includegraphics[width = 5.5cm]{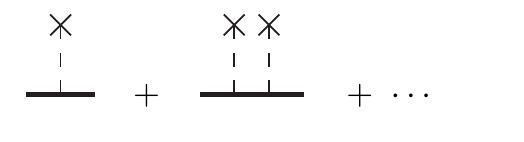}\\
b)&\\
&\includegraphics[width = 5.8cm]{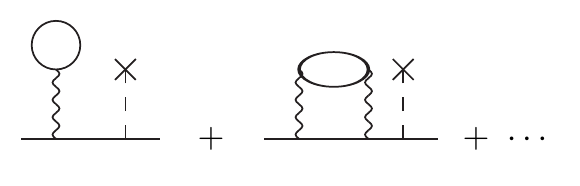}\\
c)&\\
&\includegraphics[width = 1.6cm]{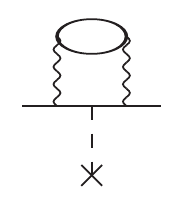}
\end{tabular}
\caption{
a) Feynman diagrams which enter the single-impurity T-matrix used in our
calculation. Bold lines are full (i.e. interacting) propagators $\hat \G^0$, the cross
denotes impurity scattering with strength $\hat g_s$.
b) Sample diagrams contained in a), where thin lines are now non-interacting propagators,
and the wavy line represents the interaction $U$.
c) An interference process between $\hat g_s$ and $U$ which is neglected in a).
}
\label{imp_dgr}
\end{figure}

In our calculations for the interacting system, we shall employ the following
approximations:
(i) We neglect interference process between different impurities, formally $\hat T_{ij}
\rightarrow \hat t \delta_{ij}$.
(ii) We capture the scattering effects using Eqs. \eqref{gf_t_matrix} and
\eqref{1_t_matrix}, i.e., we evaluate a single-impurity T-matrix $\hat t$ according to
\eqref{1_t_matrix} with full (i.e. interacting) propagators $\hat \G^0$. This neglects
processes where impurity scattering and electron-electron interactions interfere, as
illustrated in Fig.~\ref{imp_dgr}.
The change in local density of states induced by the potential scattering can then be
expressed as
\bea
\label{delta_density}
&&\delta \rho(\br, \om) \equiv \rho(\br, \om)-\rho_0(\br, \om) \\
&&= -\frac{1}{\pi}\im \mbox{Tr} \left [ \sum_i \hat \G^0(\br - {\bf R}_i, \om) {\hat
t}(\om ) \hat \G^0({\bf R}_j-\br, \om )\right ] \nonumber.
\eea
(iii) Unless otherwise noted, we employ the lowest-order Born approximation, i.e.,
$\hat t = \hat g_s$.

The approximations (i)--(iii) are designed to capture the effect of weak impurity
scattering. They do not account for impurity-induced local changes in the Kondo
screening,\cite{kaul07,morr11} a description of which would require a solution of
real-space DMFT equations with a self-consistent {\em set} of impurity
problems\cite{helmes} -- a task which we leave for future work.


\subsection{Quasiparticle interference}
\label{subsec:qpi}

If we Fourier transform $\delta \rho(\br, \om)$ \eqref{delta_density} into momentum space
and restrict ourselves to the lowest Born approximation, we find
\beq
\delta \rho(\bq, \om) = \frac{-1}{\pi}  (\sum_{j}e^{i \bq \cdot {R}_j}) \im \sum_{\bk}\left [ \hat \G^0(\bk, \om) \hat g_s \hat \G^0(\bk+\bq, \om)\right ]
\label{eq:QPI0}
\eeq
i.e, to leading order in the impurity strength, the FT of the LDOS separates into a
factor from the QPI of $\hat \G^0$, describing ``band structure'', and another factor
representing the spatial distribution of the scatters. The latter is a smooth function of
momentum; it is relevant for the overall intensity distribution of $\delta \rho(\bq, \om)$ in
momentum space, but not for the presence of sharp features (peaks, ripples etc.).

Therefore, we may restrict the calculation to the case of a single impurity, say at ${\bf
R_i=0}$. We note that this type of approximation (single impurity, T-matrix approximation
in the presence of interactions) has been frequently used to describe QPI phenomena in
various physical systems in the past (see e.g.
Refs.~\onlinecite{qpi_cup,wanglee03,qpi_pnic,qpi_topo}).
Using \eqref{tunneling_conductance}, the impurity-induced change in differential
tunneling conductance, i.e. its spatially inhomogeneous piece, is given
\bea
\delta\frac{dI}{dV}(\br, eV)&=& - g_s \rho_{tip} \frac{2 e}{\hbar} \int  d \om \,\, f^\prime (\om-eV)\times \nn
 && \hspace{-2cm} \im  \left[ t_c^2 \G^0_{cc}(\br, \om)\G^0_{cc}(-\br, \om) + t_f^2 \G^0_{fc}(\br, \om)\G^0_{cf}(-\br, \om) \right . \nn
 &&\hspace{-2cm}+ t_c t_f  \G^0_{cc}(\br, \om)\G^0_{cf}(-\br, \om) + \left . t_c t_f  \G^0_{fc}(\br, \om)\G^0_{cc}(-\br, \om)\right ].\nn
\eea
Note that no $\G^0_{ff}$ appears here because of the structure of the scattering
potential $\hat g_s$ \eqref{gs}. Transforming to Fourier space and assuming inversion
symmetry, we obtain
\bea
\delta \frac{dI}{dV}(\bq, eV)\propto   \sum_{\bk}& \im & \left [t_c^2 \,\, \G^0_{cc}(\bk, eV)\G^0_{cc}(\bk+\bq, eV)  \right . \nn
& +& \left .  t_f^2 \,\,  \G^0_{fc}(\bk, eV)\G^0_{cf}(\bk+\bq, eV) \right . \nn
  & + & \left .  2 t_c t_f \,\,  \G^0_{cc}(\bk, eV)\G^0_{cf}(\bk+\bq, eV)\right ]\nn
\label{eq:QPI}
\eea
which is the equation we will use below to generate numerical results describing QPI.
Note that the Fourier-transformed LDOS is complex in general, but the present
situation obeys inversion symmetry w.r.t. the position of the single impurity, and hence
$dI/dV(\bq, eV)$ is real.

In general, QPI phenomena as captured, e.g., by equations
\eqref{delta_density}--\eqref{eq:QPI}, lead to energy-dependent Friedel oscillations in
the LDOS, caused by scattering off the impurities. Those oscillations at an energy $\om$
are primarily determined by the shape of the iso-energy surface at $\om$ of the
underlying band structure, i.e., the oscillation wavevectors $\bq$ are given by
approximate nesting wavevectors or by wavevectors which connect points of high DOS in
momentum space. (Note that this argument neglects the influence of the real parts of the
propagators which enter in Eqs. \eqref{eq:QPI0}--\eqref{eq:QPI} as well.) Thus, the
Fourier transform of the LDOS ``ripples''\cite{qpi_cup} observed in SI-STM experiments
allows to approximately extract characteristic wavevectors $\bq(\om)$ of the underlying
electronic state. Below we shall show that QPI in heavy-fermion systems is capable of detecting
the Fermi-surface reconstruction from high to low temperatures.


\section{Energy scales and general considerations}
\label{sec:scales}

Before diving into numerical results, let us discuss a few general aspects of
heavy-fermion physics relevant to SI-STM experiments performed at variable temperature.

Provided that the Anderson model \eqref{PAM} is close to the Kondo limit, i.e., $U,
-\ep_f \gg V^2/W$ where $W$ is the half-bandwidth of the $c$ electrons, local moments
will form on the $f$ sites upon cooling below a temperature of order $U$. At those high
temperatures, conduction electrons will weakly scatter off those local moments, a process
which can be described in perturbation theory in the Kondo coupling $\JK$ in \eqref{KLM}.
Upon cooling, the scattering intensity becomes large when $T$ reaches the single-impurity Kondo
temperature $\TK$ where perturbation theory breaks down.

In the opposite limit of low
temperatures (and in the absence of spontaneous symmetry breaking e.g. by magnetism or
superconductivity), a heavy Fermi liquid forms away from half-filling. This Fermi liquid
is adiabatically connected to the weakly interacting limit of the PAM \eqref{PAM}, and
perturbation theory in $U$ is at least formally applicable. The Fermi liquid is bounded
in temperature from above by a coherence temperature $\Tcoh$ which acts as an effective
Fermi energy for the heavy quasiparticles. Typically $\Tcoh<\TK$.

In the low-temperature Fermi-liquid regime, Luttinger's theorem applies, and consequently
the Fermi volume is ``large'', i.e., given by $\mathcal V_{\rm FL}=K_d (n_{\rm tot}\,\mbox{mod} 2)$, where
$n_{\rm tot}=n_c+n_f$ is the total number of electrons per unit cell and $K_d=(2\pi)^d/(2
v_0)$ a phase space factor, with $v_0$ the unit cell volume and $d$ the spatial
dimensionality. In the limit of small $\TK$, it also makes sense to consider a Fermi
volume for temperatures $\TK \ll T \ll W$. Here, the $f$ electrons do not contribute,
leading to a ``small'' Fermi volume with $\mathcal V_{s}=K_d n_c$.

Inelastic scattering between $c$ and $f$ electrons is particularly strong for
intermediate temperatures near $\Tcoh,\TK$. In fact, a common experimental definition of
$\Tcoh$ is via the maximum in the resistivity $\rho(T)$. Thus, one expects that the
temperature-driven crossover between the small and large Fermi volumes occurs at a
temperature of order $\Tcoh,\TK$.

\begin{figure}[!t]
\centering
\includegraphics[width = 6.8cm]{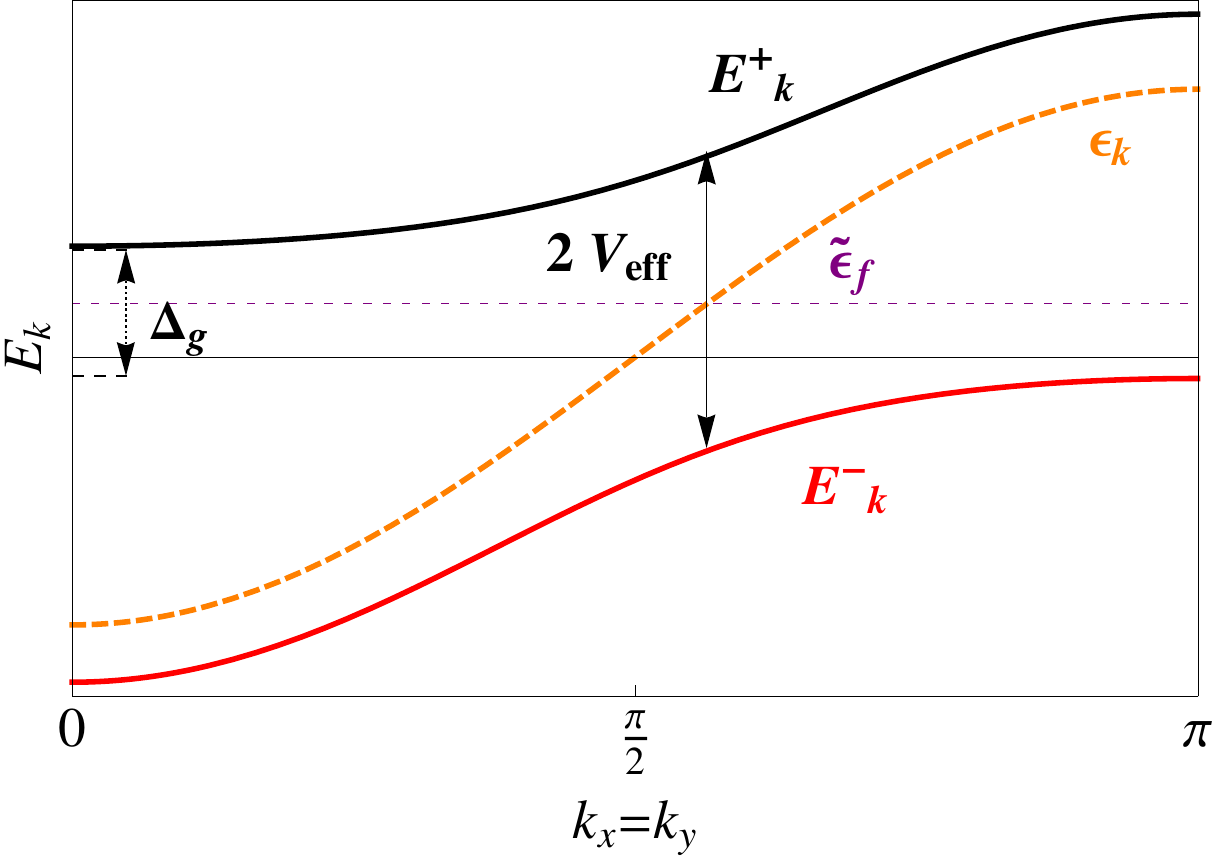}
\caption{
Schematic representation of the $T\rightarrow 0$ band structure expected for a PAM.
$V_{\rm eff}$ and $\tilde \ep_f$ are the renormalized hybridization and f-chemical potential
respectively. $\Delta_g$ is the hybridization gap between the lower and upper
renormalized bands.
}
\label{bands}
\end{figure}

The physics of the heavy Fermi liquid can be understood, to some degree, in a picture of
renormalized bands: Two bands of non-interacting $c$ and $f$ fermions
hybridize with a renormalized hybridization $V_{\rm eff}$. This results in two
quasiparticle bands with dispersions
\beq
E_{\bk}^\pm=\frac{\ek+\tilde \ep_f \pm \sqrt{(\ek-\tilde \ep_f)^2+4V_{\rm eff}^2}}{2},
\label{renormalized_bands}
\eeq
where $\tilde \ep_f$ is the renormalized energy of the $f$-electrons.  These bands
describe sharp quasiparticles formed as a mixture of $f$ and $c$ degrees of freedom. The
renormalized bands, shown schematically in Fig.~\ref{bands}, are separated
in energy by a direct (or optical) gap $\Delta_{\rm opt} = 2V_{\rm eff}$ at the crossing of the
original conduction band and $\tilde \ep_f$, while the bottom of the upper band is
separated from the top of the lower band by an indirect gap $\Delta_g \propto
\sqrt{V_{\rm eff} W}$. $\Delta_g$ is typically called hybridization gap.

%
%

While this two-band picture is trivially realized at $U=0$ in \eqref{PAM}, with $V_{\rm
eff}=V$ and $\tilde \ep_f = \ep_f$, it can be obtained using a slave-boson mean-field
approximation from either Eq.~\eqref{PAM} in the $U\rightarrow \infty$ limit or from
Eq.~\eqref{KLM}. In both cases, a slave boson $b_i$ is introduced such that
\beq
\fda_{i, \si} \rightarrow \tilde{f}^\dagger_{i, \si} b_i.
\label{slave_boson_transformation}
\eeq
In the simplest approximation, $b$ is uniform and static and measures the effective
hybridization between the bands, i.e., $V_{\rm eff}$ in Eq.~\eqref{renormalized_bands}
becomes $V_{\rm eff}=b V$ ($V_{\rm eff}=b \JK$) in the Anderson (Kondo) model.
(Note that $b$ is also a measure of the mass renormalization of the quasiparticles, i.e.,
the mass enhancement is given by $m^\ast/m = 1/b^2$.)
This mean-field approximation can be formally justified in a limit where the spin symmetry of
the original model is extended from SU(2) to SU($N$) with
$N\rightarrow\infty$.
Remarkably, the qualitative picture of effective quasiparticle bands, Fig.~\ref{bands},
has been reproduced in numerical treatments of the periodic Anderson (or Kondo lattice)
models, using both DMFT \cite{grenzebach06,assaad_DMFT} and its cluster generalizations
\cite{assaad_CDMFT}. Hence, a quasiparticle description of low-temperature heavy-fermion
bands appears justified, with a number of caveats noted in the following.

While the slave-boson mean-field theory correctly captures the exponential dependence of
the Kondo scale $\TK$ on $\JK$, $W$, and accounts for the enlargement of the Fermi
surface in the Fermi-liquid regime, it misses all inelastic scattering processes, as it
only operates with non-interacting fermions. Consequently, the physics not only at finite
temperatures, but also that at $T\to 0$ and finite energies is not described well. Both
interaction-induced broadening of spectral features away from the Fermi level and incoherent scattering
at elevated
temperatures are not included. For instance, the hybridization gap in the
heavy-fermion state, Fig.~\ref{bands}, is predicted to be a sharp gap in the mean-field
theory, but can be expected to be significantly smeared or even washed out by interaction
effects. In fact, it has been argued that the experimentally observed Fano-line shape
is a consequence of a quasiparticle broadening not captured by the mean-field
theory.\cite{coleman09, woelfle10}
More seriously, the crossover from $T\ll\TK$ to $T\gg\TK$ turns into an
artificial phase transition at $\TK$ in the slave-boson theory. Hence, to understand the
crossover between large and small Fermi volumes upon varying temperature requires at least to treat local correlation
effects in a non-perturbative manner -- this is what we shall do below using DMFT.


\section{Numerical results}
\label{sec:res}

We have performed extensive studies of the paramagnetic phase of the PAM using DMFT-NRG at various
temperatures, tracking the formation of the heavy Fermi liquid upon cooling out of
the light conduction band and the $f$-electrons.

For simplicity, we consider a square lattice with nearest-neighbor hopping,
\begin{equation}
\ek = - 2 t (\cos k_x + \cos k_y) + \ep_c \,,
\label{eq:disp}
\end{equation}
and half-bandwidth $W=4t$ which we employ as energy unit. We assume the chemical
potential to be located at zero energy, such that $\ep_c$ controls the conduction-band
filling.

Most of the figures shown have been obtained with the following set of the PAM
parameters: $\ep_c=0.4$, $\ep_f=-0.9$, $V=0.55$, $U=10$ (in units of $W$). With these
parameters the Anderson width is $\Gamma_0 \approx 0.51$. In the $T\rightarrow0$ limit,
we find the occupation number of the $c$ ($f$) electrons is $n_c\approx 0.57$ ($n_f
\approx 0.87$), i.e., we work close to the Kondo regime, with a deviation from integer
filling comparable to that in actual Ce or Yb heavy-fermion systems. The temperature
variation of the band fillings is less than 5\% within the temperature range considered
here.
For the above parameters, we find a $T\to 0$ mass renormalization
\begin{equation}
m^\ast/m = \left . 1 - \frac{\partial \re\,\Sigma_f(\om)}{\partial \om}\right|_{\om=0} \approx 60
\end{equation}
which allows to define a low-temperature Kondo scale $\bar\TK = \Gamma_0\, m/m^\ast
\approx 8 \times 10^{-3}$ -- this roughly matches the full width at half maximum (FWHM)
of the $T=0$ Abrikosov-Suhl resonance in $\G_{ff}$, see Fig.~\ref{results:spectral}
below. Note, however, that a unique definition of neither the Kondo nor the coherence
scale exists, and depending on ones choice may differ by up to an order of magnitude.
In our case, we find the temperature scale where the impurity entropy within DMFT reaches $0.5\ln 2$
to be $T_{1/2} \approx 1.5 \times 10^{-3}$, and heavy-fermion bands are fully formed only below
a {\em coherence temperature} of $\Tcoh \approx 2 \times 10^{-4}$.

For the 2d system considered here, the van-Hove singularity (vHs) arising
from the saddle point in the dispersion is a prominent feature in the DOS which will play
a role in all momentum-integrated spectra discussed below.


\begin{figure}[!t]
\centering
\includegraphics[width=5.6cm]{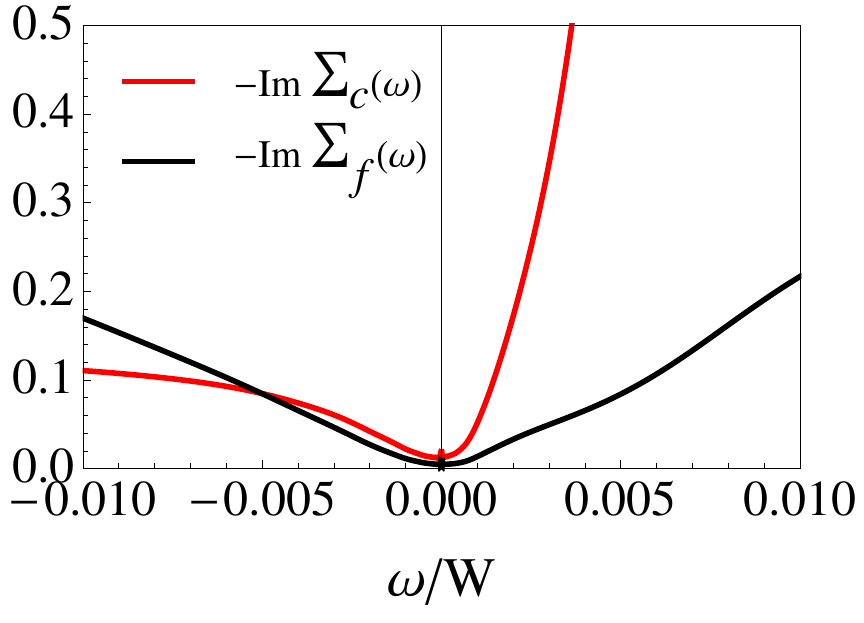}
\caption{
Frequency dependence of the self-energies, $\im \Sigma_f$ and $\im \Sigma_c$, of the $f$
and $c$ electrons, respectively, calculated using DMFT-NRG for the periodic Anderson
model with parameter $U=10, \ep_f=-0.9, \ep_c=0.4, V=0.55$ at a low temperature
of $T=1.8\times10^{-6}$. Energies are given in units of the half-bandwidth $W$.
}
\label{results:Sff_Scc}
\end{figure}

\subsection{Local electronic spectra and self-energies}
\label{subsec:spectral_functions}

Once we have solved the self-consistent equation \eqref{PAM_self_consistency} for the $f$
self-energy $\Sigma_f$, we can form the Green's function matrix whose elements are given
in (\ref{greens_cc}--\ref{greens_matrix}). In this section we discuss self-energies as
well as the local spectral functions the $c$ and $f$ electrons at different temperatures.

\subsubsection{Self-energies}

As a check of Fermi-liquid behavior, we start with the frequency dependence of the
imaginary parts of the low-temperature electronic self-energies, $\im \Sigma_c$ and $\im
\Sigma_f$, displayed in Fig.~\ref{results:Sff_Scc}. Both exhibit quadratic behavior in
$\om$, in a limited energy range around the Fermi level, as expected for a Fermi liquid.
We note that the self-energies have a small residual value at $\om=0$, arising from both
inaccuracies of NRG in fulfilling spectral sum rules\cite{NRGrev} and the artificial
broadening which has been introduced to stabilize the numerics.\cite{grenzebach06} (Our
results do not depend on details of this broadening.)

\subsubsection{Low-temperature spectra}

\begin{figure}[!t]
\centering
\begin{tabular}{c}
\includegraphics[width = 6.4cm]{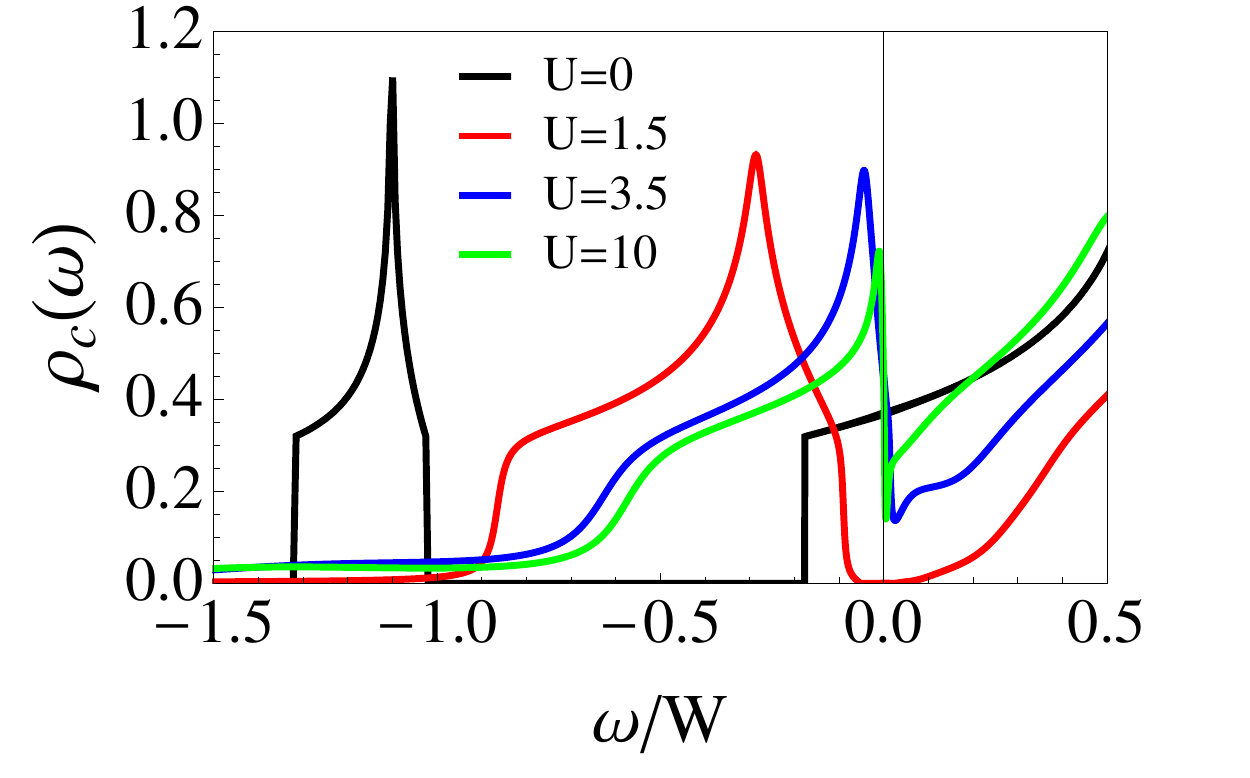} \\
\hspace{-0.5cm}\includegraphics[width = 5.8cm]{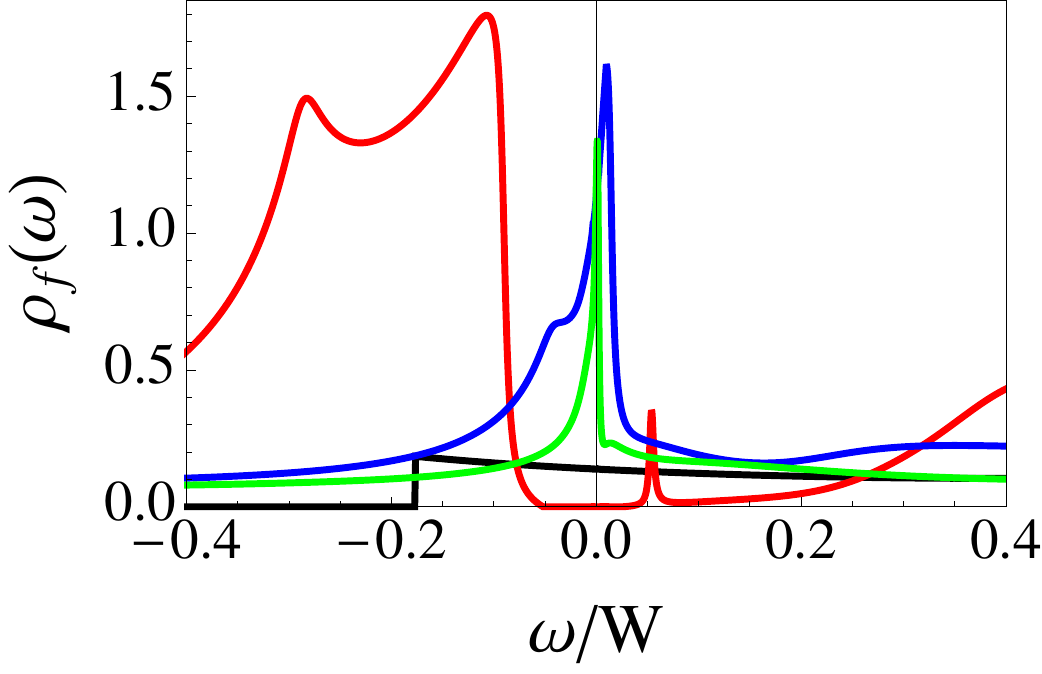}
\end{tabular}
\caption{
a) $c$ and b) $f$ electron local spectral functions of the periodic Anderson model,
calculated using DMFT-NRG at $T=1.8\times10^{-6}$. The different curves correspond to
different values of the local interaction $U$; other parameters are as in Fig.~\ref{results:Sff_Scc}.
Because the chemical potential is kept fixed, the total and individual band fillings vary as follows:
$n_{\rm tot}-n_c-n_f \approx 2.2-0.42-1.78$  ($U=0$), $1.98-0.78-1.2$ ($U=1.5$), $1.52-0.63-0.89$ ($U=3.5$), and $1.44-0.57-0.87$ ($U=10$).
}
\label{results:spectral_U}
\end{figure}

%
\begin{figure*}[!tbh]
\centering
\includegraphics[width=17cm]{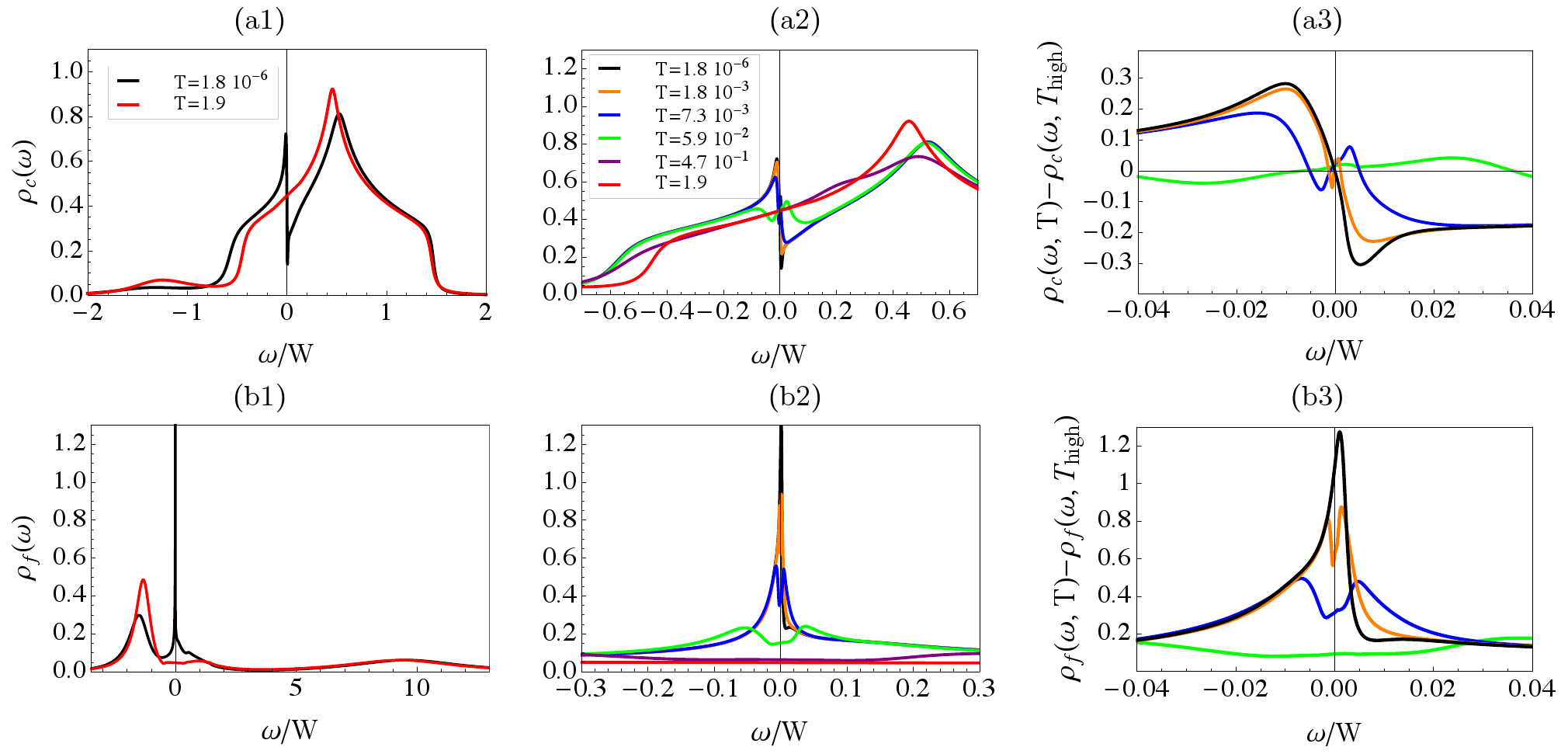}
\caption{
Local spectral functions a) $\rho_c(\om)$ and b) $\rho_f(\om)$ of the PAM
calculated using DMFT-NRG for different temperatures $T$, with model parameters as in
Fig.~\ref{results:Sff_Scc}.
a1,b1) Spectra on the full energy range for two temperatures, $T \gg \TK$ and $T \ll
\TK$.
a2,b2) Temperature evolution of $\rho_c$ and $\rho_f$ shown in a restricted energy range
near the Fermi level.
a3,c3) Low-temperature, low-energy results for $\rho_c$ and $\rho_f$, with the
high-temperature spectra at $T=4.7 \times 10^{-1}$ subtracted. The progressive formation
of the hybridization pseudogap (panel a) and the Abrikosov-Suhl resonance (panel b) are
clearly visible.
}
\label{results:spectral}
\end{figure*}

Next we turn to the local spectral functions, $\rho_c(\om) =(-1/\pi)\im \G_{cc}(\om)$ and
$\rho_f(\om) =(-1/\pi)\im \G_{ff}(\om)$. To illustrate the effect of interactions and the
concomitant deviations from the two-band picture of slave-boson theory explained in
Sec.~\ref{sec:scales}, we show the $U$ dependence of the low-temperature spectra in
Fig.~\ref{results:spectral_U}.
For the $c$ spectral function, Fig.~\ref{results:spectral_U}a, increasing $U$ has the
following effects: a renormalization of the band positions arising from the real part of
the self-energy (note that the chemical potential is kept fixed), a renormalization and
simultaneous displacement of the inter-band gap, and its smearing due to the
quasiparticle broadening induced by $\im \Sigma_c(\om)$. While the two first effects are
accounted for in the slave-boson theory, the last is absent in this approximation.
The progressive smearing of the hard gap at $U=0$ and its displacement are also visible
in the $f$ spectrum, Fig. \ref{results:spectral_U}b. Increasing $U$ induces the formation
of the Kondo peak (or Abrikosov-Suhl resonance) near the Fermi level. Its width can be
considered as a measure of the Kondo scale $\TK$, which approaches a finite value in the
$U \rightarrow \infty$ limit (due to the finite $\ep_f$).

A few comments are in order: A sharply defined hybridization gap in $\rho_c$,
corresponding to the indirect gap $\Delta_g$ of Fig.~\ref{bands}, is only present for
$U\lesssim 2$, while it is reduced to a dip or pseudogap for larger values of $U\gtrsim
3$. Remarkably, with increasing $U$ one observes an apparent pinning of the vHs of the
$c$ band to the Fermi level. As will become clear from the band structures discussed
below, this is due to the progressive narrowing of the lower quasiparticle band at low
energies. As a result, the vHs and hybridization gap feature cannot be clearly separated
in $\rho_c$ (both exist on an energy scale of order $\TK$). The vHs is also responsible
for the strong particle-hole asymmetry of the $c$ spectra.
In $\rho_f$ both the vHs and the hybridization gap are less prominent. Finally, we remind
the reader that our single-orbital PAM does not account for multiple $f$ levels and their
crystal-field splitting, i.e., we only model the physics of the lowest-energy
crystal-field doublet.

\begin{figure*}[!bth]
\centering
\includegraphics[width = 17.5cm]{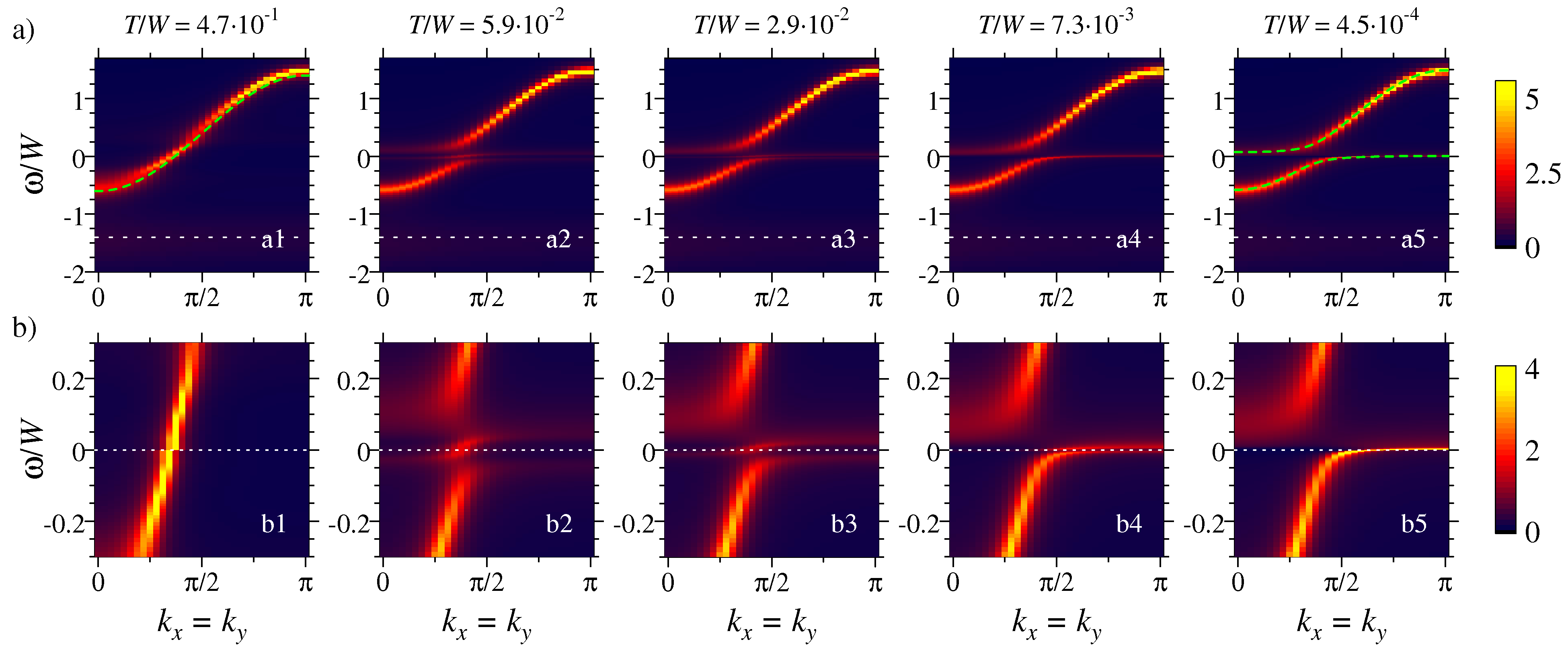}
\caption{
Evolution of the electronic spectrum of the PAM, with parameters as in
Fig.~\ref{results:Sff_Scc}, with temperature: Shown is $(-1/\pi)\im (\G_{ff} + \G_{cc})(\bk,\om)$
as function of $k_x=k_y$ and $\om$ over a) the full range of energies and b) a restricted
energy window near the Fermi level.
In a) the dotted horizontal lines mark the renormalized $f$ level (i.e. the lower Hubbard
band), while in b) the dotted lines represent the Fermi level. Further the dashed line in
a1) shows the bare conduction band $\ek$. At low temperatures, the formation of heavy
bands is obvious, with the reconstruction happening mainly within the optical gap. In a5)
the dashed lines represent a fit of the intensity at elevated energies to the effective two-band
picture using \eqref{renormalized_bands} with $V_{\rm eff}=0.17, \tilde \ep_f=0.025$, and
$\tilde \ep_c = 0.47$, for detail see text.
}
\label{results:renormalized_bands}
\end{figure*}

\subsubsection{Temperature evolution of spectra}

The temperature evolution of both $c$ and $f$ local spectral functions in the Kondo
regime ($U=10$) is shown in Fig. \ref{results:spectral}. As can be seen in
Fig.~\ref{results:spectral}b1, the $f$ spectrum has the well-known three-peak structure at low
temperature: the two charge excitation peaks at $\ep_f$ and $\ep_f+U$, well separated
from the Abrikosov-Suhl resonance at the Fermi level. The latter is absent at high
temperature and is the fingerprint of the Kondo effect. A close-up view of $\rho_f(\om)$
near the Fermi level, Figs.~\ref{results:spectral}b2 and \ref{results:spectral}b3, shows
the development of this resonance with decreasing temperatures.

In the local $c$ spectrum, Fig. \ref{results:spectral}a1, cooling induces the
hybridization pseudogap and a second van-Hove singularity near the Fermi level. This is
again shown in detail in Figs.~\ref{results:spectral}a2 and \ref{results:spectral}a3,
where the development of a positive-frequency dip with decreasing temperatures is
emphasized. As noted above, this dip is what remains of the hybridization gap once
inelastic scattering is fully taken into account; its particle-hole asymmetry is a
band-structure effect arising from the low-energy vHs in the present 2d case.
The displacement of the higher-energy vHs in Fig.~\ref{results:spectral}a2
can be taken as a measure of $\re \Sigma_c$. The low-energy structures apparent at
intermediate temperature $\approx 10^{-3} \ldots 10^{-2}$ are connected with the
dynamical character of the band reconstruction and will be discussed in more detail
below.

\subsection{Renormalized band structure}
\label{subsec:bands}

After presenting the local (i.e. momentum-integrated) spectral functions, we turn to the
full momentum dependence of the spectra. (Recall that in DMFT the interaction-induced
self-energies are momentum-independent, but the Green's functions acquire momentum
dependence from the non-interacting bands.)
Fig.~\ref{results:renormalized_bands} illustrates the temperature evolution of the electronic
spectrum in a false-color plot of $(-1/\pi) \im (\G_{ff} + \G_{cc})(\bk,\om)$ as function of energy
$\om$ and momentum $\bk$ along the line $k_x=k_y$.
At high temperature $T\gg \TK$, Fig.~\ref{results:renormalized_bands}a1, the signal
consists of the rather sharp bare $c$ band and the ``atomic'' $f$ levels at $\ep_f$ and
$\ep_f+U$, the latter being broadened by the hybridization. In the opposite limit $T \ll T_K$,
Fig.~\ref{results:renormalized_bands}a5, we have two self-energy-broadened bands, both
being rather flat near the Fermi level. One may fit this low-temperature band structure
using the two-band picture described in Sec.~\ref{sec:scales}. However, we find that
there is {\em no} unique fit for all energies: The features at elevated energies are well
described by the fit shown in Fig.~\ref{results:renormalized_bands}a5, which, however,
overestimates the low-energy slope of the heavy band by a factor of 2.5. Alternatively, a
low-energy fit results in $V_{\rm eff}=0.072$, consistent with $m^\ast/m \approx 60$
obtained from DMFT self-energy -- this implies that the DMFT result has to be understood
in terms of an energy-dependent band hybridization.

The reconstruction of the band structure from a weakly interacting $c$ band and $f$
levels at high temperature to two heavy bands at low temperature occurs mainly within the
optical gap $\sim 2 V_{\rm eff}$, as can be seen from Fig.~\ref{results:renormalized_bands}b:
Portions outside this window show little variation with temperature. When temperature is
decreased, spectral weight is transferred gradually from the bare $c$ band near the small
Fermi surface to the parts of the incipient heavy-fermion bands near the boundaries of the
momentum window. Importantly, for a given energy this weight transfer happens at an
energy-dependent temperature, i.e., upon cooling higher energies are reconstructed first,
Figs.~\ref{results:renormalized_bands}b2--\ref{results:renormalized_bands}b4. As a result
of this {\em dynamical} reconstruction, there is an ``island'' of spectral intensity near
the bare $c$ band left at intermediate temperature,
Fig.~\ref{results:renormalized_bands}b2, which shrinks and finally disappears upon
cooling to significantly lower $T$.

%
%
\begin{figure*}[!t]
\centering
\includegraphics[width = 18cm]{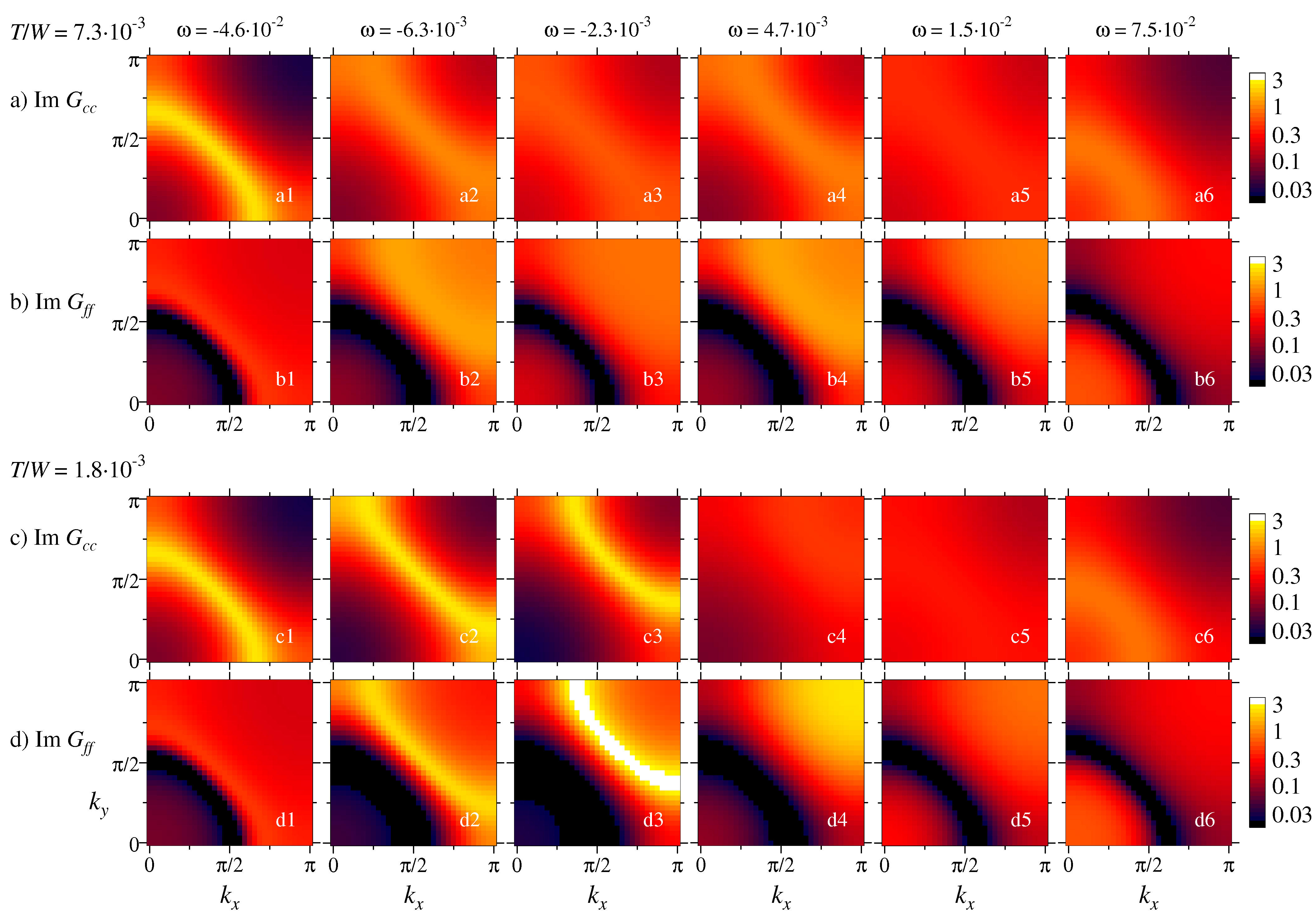}
\caption{
Constant-energy cuts through the electronic spectra of the PAM, with parameters as in
Fig.~\ref{results:Sff_Scc}. Panels a) and c) show $(-1/\pi)\im \G_{cc}(\bk,\om)$ in a
quarter of the first Brillouin zone, panels b) and d) show $(-1/\pi) \im \G_{ff}(\bk,\om)$.
The temperatures are a,b) $T=7.3 \times 10^{-3}$ and c,d) $T=1.8 \times 10^{-3}$.
Note that the intensity color scale is logarithmic in all cases.
}
\label{results:spec_om}
\end{figure*}

Constant-energy cuts of the spectral functions, now separated into $(-1/\pi)\im \G_{ff}$
and $(-1/\pi)\im \G_{cc}$, illustrating the weight distribution in 2d
momentum space, are shown in Fig.~\ref{results:spec_om} for two different temperatures
(higher than the one at which the renormalized bands are fully constructed, which is
roughly $2\times 10^{-4}$). The plots allow to track in detail the spectral-weight
transfer from the high-temperature bare conduction band to the low-temperature
renormalized bands; note that we have employed a logarithmic color scale in order to
visualized weak-intensity features as well.

Starting the discussion with $c$ electron spectrum, we confirm that there is essentially
no temperature dependence for energies outside the optical gap, compare panels a1) and
a6) to c1) and c6) in Fig.~\ref{results:spec_om}. In contrast, for energies inside the
optical gap one observes a significant temperature dependence: At higher temperatures a
rather diffuse signal is present in panels a2)--a5), while quasiparticle peaks form
at low $T$ and negative energies, panels c2) and c3), and spectral weight disappears
at low $T$ and positives energies, panels c4) and c5) -- the latter correspond to
energies inside the hybridization gap. Note that panel Fig.~\ref{results:spec_om}a4
contains shows a well-defined (albeit weak) iso-energy contour -- this
represents a yet unreconstructed piece of the light band which only disappears at lower
temperature, see also Fig.~\ref{results:renormalized_bands}b.

The $f$ electron spectra in all panels, Fig.~\ref{results:spec_om}b,d, show a
pronounced weight {\em reduction} which follows closely the ``small'' Fermi surface
defined by $\om-\epsilon_\bk -\re \Sigma_c(\om) =0$; this effect can be easily deduced
from Eq.~\eqref{greens_ff} (given that the interaction-induced self-energy has a non-zero
imaginary part). A well-defined quasiparticle peak is only visible in $\G_{ff}$ at
frequencies where portions of the new renormalized bands are constructed, i.e., in panels
d2) and d3).


\subsection{Differential tunneling conductance}
\label{subsec:dtc}

In the presence of lattice translational symmetry, i.e., in the absence of impurities,
the differential tunneling conductance $dI/dV(\br_i, \om)$,
Eq.~\eqref{tunneling_conductance}, is independent of the spatial position (i.e. lattice
site) $\br_i$. It is a measure of the local density of states, but depends on the
ratio $t_f/t_c$ of the tunneling amplitudes.

\begin{figure}[!t]
\centering
\includegraphics[width = 8.7cm]{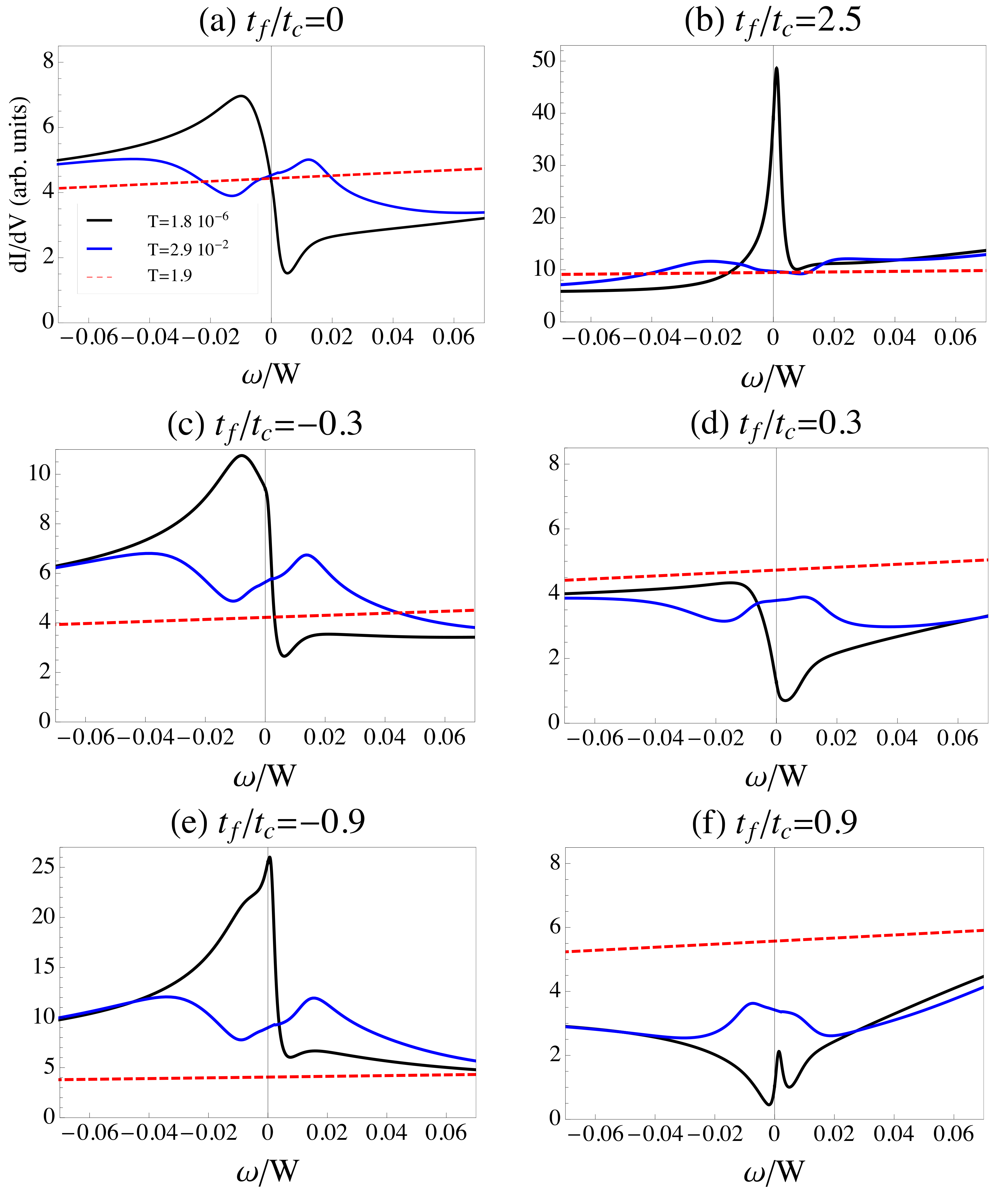}
\caption{
Evolution of the differential conductance $dI/dV$ for small bias voltage with the ratio
$t_f/t_c$ for different temperatures $T\gg T_K$, $T\sim T_K$ and $T\ll T_K$. The effect
of destructive [constructive] interference for positive [negative] ratios $t_f/t_c$ is
visible at low temperatures in panels b), d), f) [c) and e)], while the interference is
essentially ineffective at high $T$.
}
\label{results:tunneling_conductance}
\end{figure}

Fig.~\ref{results:tunneling_conductance} shows our results for the energy (or
bias-voltage) dependence of $dI/dV$ for different values of $t_f/t_c$ and each for three
representative temperatures, $T\gg T_K$, $T\sim T_K$, and $T \ll T_K$.
For small values of $t_f/t_c$, signatures of the hybridization gap are clearly seen, but
 -- as expected -- there is never a hard gap near the Fermi level, in contrast to
results from slave-boson mean-field theory.\cite{coleman09,morr10,woelfle10} The reason
of course are inelastic scattering processes, captured by DMFT but absent in the static
mean-field approximation. (Those were added phenomenologically to explain the smearing of
the hard gap near the Fermi level and the appearance of the asymmetric peak in
experiments.\cite{woelfle10})

For $t_f=0$, Fig.~\ref{results:tunneling_conductance}a, the differential tunneling
conductance is proportional to the $c$-electron LDOS. It is almost flat at high
temperature while it has a broad peak and a dip at low $T$, resulting from the interplay
of the vHs and the hybridization gap. The particle--hole asymmetry is thus inherited from
the conduction band. This asymmetry appears opposite to the one observed
experimentally\cite{davis_urusi, yazdani_urusi} in URu$_2$Si$_2$; in our calculations,
such asymmetry would result upon choosing $n_c>1$. (Note that the asymmetry in experiment
may be influenced by many factors: conduction band dispersions and fillings, relevant
crystal field levels of the Kondo ion, and the tunneling paths relevant for STM.)

For $t_f/t_c \gg 1$, the main contribution to the
differential tunneling conductance comes from the $f$-electron LDOS. Accordingly, a large
(Kondo) peak is observed in $dI/dV$ near $\om=0$ and the low-energy particle-hole
asymmetry is inverted. This is shown in Fig.~\ref{results:tunneling_conductance}b.
%

%

\begin{figure*}[!t]
\centering
\includegraphics[width = 17.5cm]{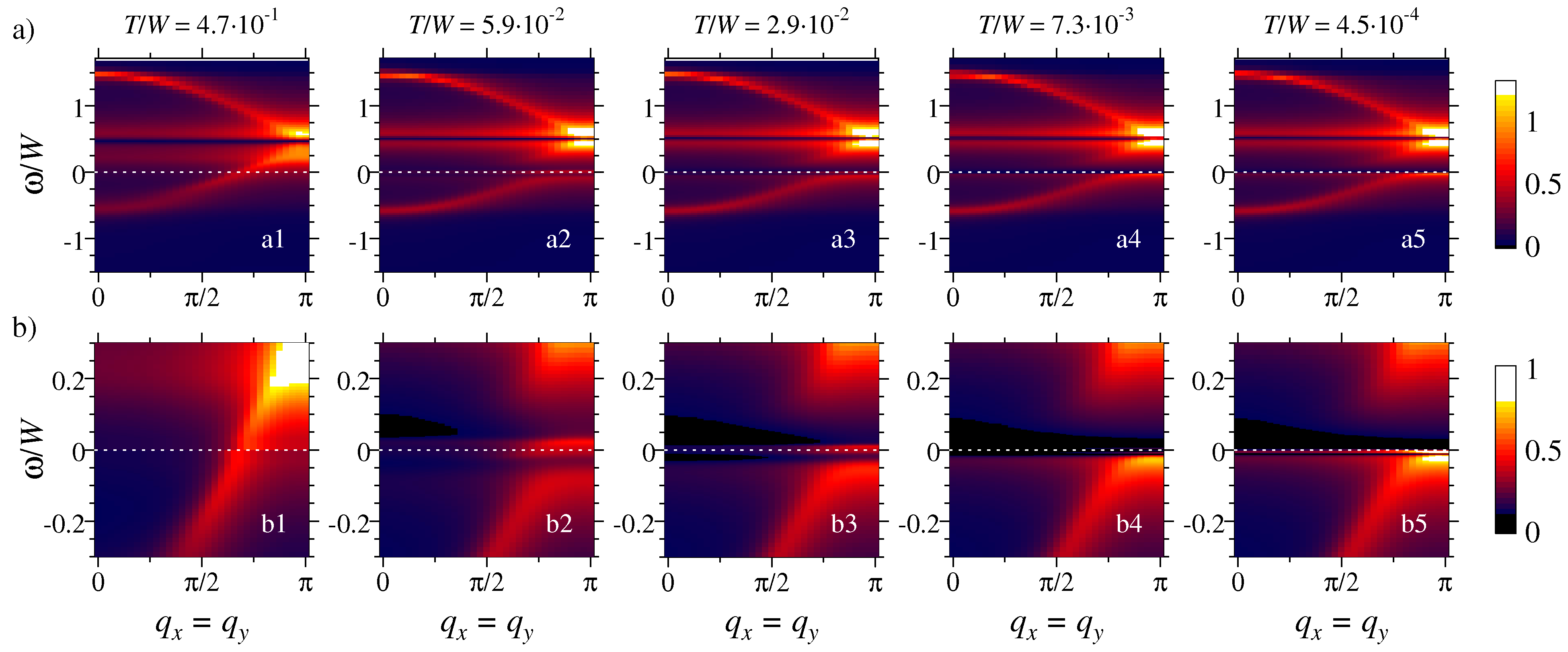}
\caption{
a) Intensity map of the Fourier transform of the spatially inhomogeneous part of the
differential tunneling conductance, $|\delta \frac{dI}{dV}(\bq, \om)|$, induced by a single
point-like weak scalar impurity (i.e. the QPI signal), along the $q_x=q_y$ direction of
the first Brillouin zone. The perculiar behavior near $\om/W=0.5$ is due to perfect
nesting of $\ek$, for details see text.
b) Close-up near the Fermi level to be compared with the single-electron spectrum in
Fig.~\ref{results:renormalized_bands}b. The model parameters are as in
Fig.~\ref{results:Sff_Scc}, the ratio of the tunneling amplitudes is $t_f/t_c=0$.
}
\label{results:QPI_E}
\end{figure*}

For intermediate values of $t_f/t_c$, quantum interferences between the two tunneling
channels have dramatic effects at low temperature as can be seen in
Fig.~\ref{results:tunneling_conductance}c--f. These effects depend on the sign of
$t_f/t_c$. With increasing positive $t_f/t_c$ and at low temperatures, the vHs peak at
negative bias gradually decreases. Then the low-energy particle--hole asymmetry in
$dI/dV$ is inverted, Fig.~\ref{results:tunneling_conductance}d,f, while the Kondo peak
emerges near $\om=0$. Notice that, while the ratio $t_f/t_c$ increases, the differential
tunneling conductance at low $T$ decreases below the one at high temperature before it
increases drastically for $t_f/t_c \gg 1$. The initial decrease is due to {\em
destructive interference} of the two tunneling channels arising from a negative $\im
\G_{cf}(\om)$ (for the $\om$ range considered here) and positive ratio $t_f/t_c$. (This
effect is essentially ineffective at high temperature.) Conversely, for negative
$t_f/t_c$ we observe {\em constructive interference} of the tunneling channels at low
$\om$ and $T$, as can be seen in Fig.~\ref{results:tunneling_conductance}c,e.
Note that the type of interference (constructive vs. destructive) depends on the sign of
$t_f t_c \im \G_{cf}(\om)$, the latter in turn depending on particle-hole
asymmetry and microscopic details.


\subsection{Quasiparticle interference}
\label{subsec:QPI}

Finally, we study quasiparticle-interference phenomena induced by scattering off
impurities. As detailed in Sec.~\ref{subsec:impurities}, we concentrate on the physics of
a single point-like scatterer in the Born limit, which should be appropriate for dilute
weak impurities. As is done experimentally, we focus on the intensity in energy-momentum
space of the Fourier-transformed differential tunneling conductance, i.e., the magnitude
of $|\delta \frac{dI}{dV}(\bq, E)|$ in Eq.~\eqref{eq:QPI}.


\subsubsection{QPI and renormalized band structure}

We start the discussion of QPI with a plot of $|\delta \frac{dI}{dV}(\bq, eV)|$ along
$q_x=q_y$ for $t_f/t_c=0$, Fig.~\ref{results:QPI_E}, to be compared to the single-particle
spectrum shown in Fig.~\ref{results:renormalized_bands}.

To understand the QPI result, one has to recall that high QPI intensity at a given energy
$\om$ is expected -- within a quasiparticle picture -- for wavevectors which connect
pieces of the quasiparticle iso-energy contour at $\om$. Thus, for a simple
inversion-symmetric dispersion $\ek$, high QPI intensity at $\om=\ek$ will occur for
$\bq=2\bk$ such that $\bq$ connects the equal-energy momenta $-\bk$ and $\bk$.

Indeed, by using $\bk=\bq/2$ and ``unfolding'' the plot of
$|\delta\frac{dI}{dV}(\bq,eV)|$ (i.e. adding a mirror image), one can qualitatively
recover from Fig.~\ref{results:QPI_E} the single-particle spectrum shown in
Fig.~\ref{results:renormalized_bands}. In particular, the reconstruction of the band
structure upon cooling, where spectral weight is transferred within the optical gap near
the Fermi level (Sec.~\ref{subsec:bands}), is clearly visible in the QPI plot as well,
Fig.~\ref{results:QPI_E}.

Of course, more complicated band structures, their nesting properties, and the effect of
van-Hove singularities are not captured by the above argument. This is nicely visible in
our data as well: The QPI signal near the middle of the band is influenced by the nesting
properties of the square-lattice dispersion, Eq.~\eqref{eq:disp}. The iso-energy contour
is a rotated square with perfect nesting in the non-interacting limit. As a result, that
QPI response is very strong at $\bq=(\pi,\pi)$, but extends to all wavevectors with
$q_x=q_y$, albeit with a reduced intensity, thanks to the interaction-induced
quasiparticle broadening, compare Fig.~\ref{results:QPI_E}a near $\om/W\approx 0.5$.


\subsubsection{Temperature evolution of QPI}

\begin{figure*}[!b]
\centering
\includegraphics[width = 17.5cm]{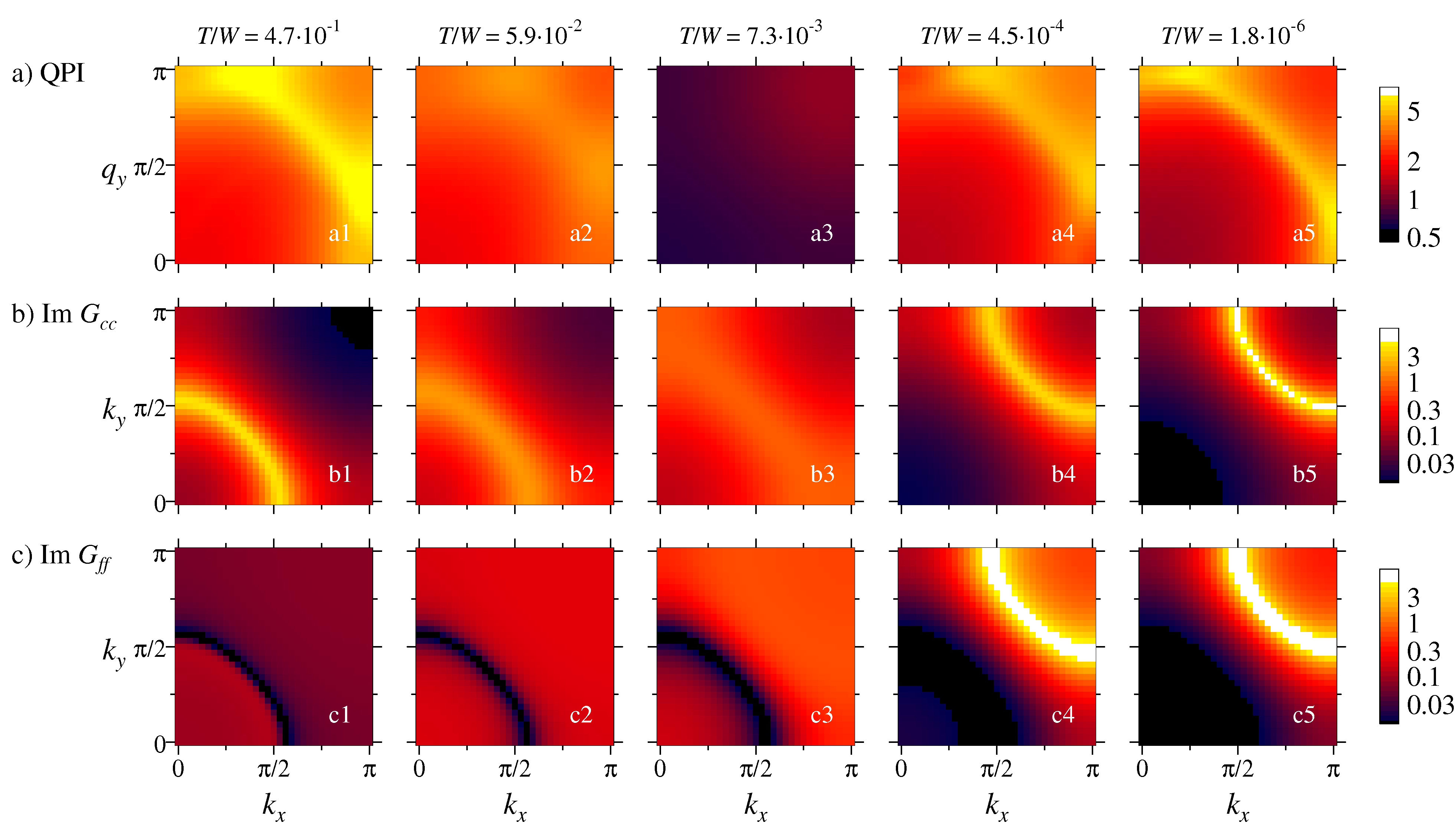}
\caption{
Constant-energy cuts through
a) the QPI signal,
b) the $c$-electron spectrum, and
c) the $f$-electron spectrum, showing a quarter of the first Brillouin zone for
different temperatures at an energy $\om=0$. The ratio of the tunneling amplitudes is
$t_f/t_c=0$.
The model parameters are the same as in Fig.~\ref{results:Sff_Scc}.
}
\label{qpi_fermi}
\end{figure*}

\begin{figure*}[!t]
\centering
\includegraphics[width = 17.5cm]{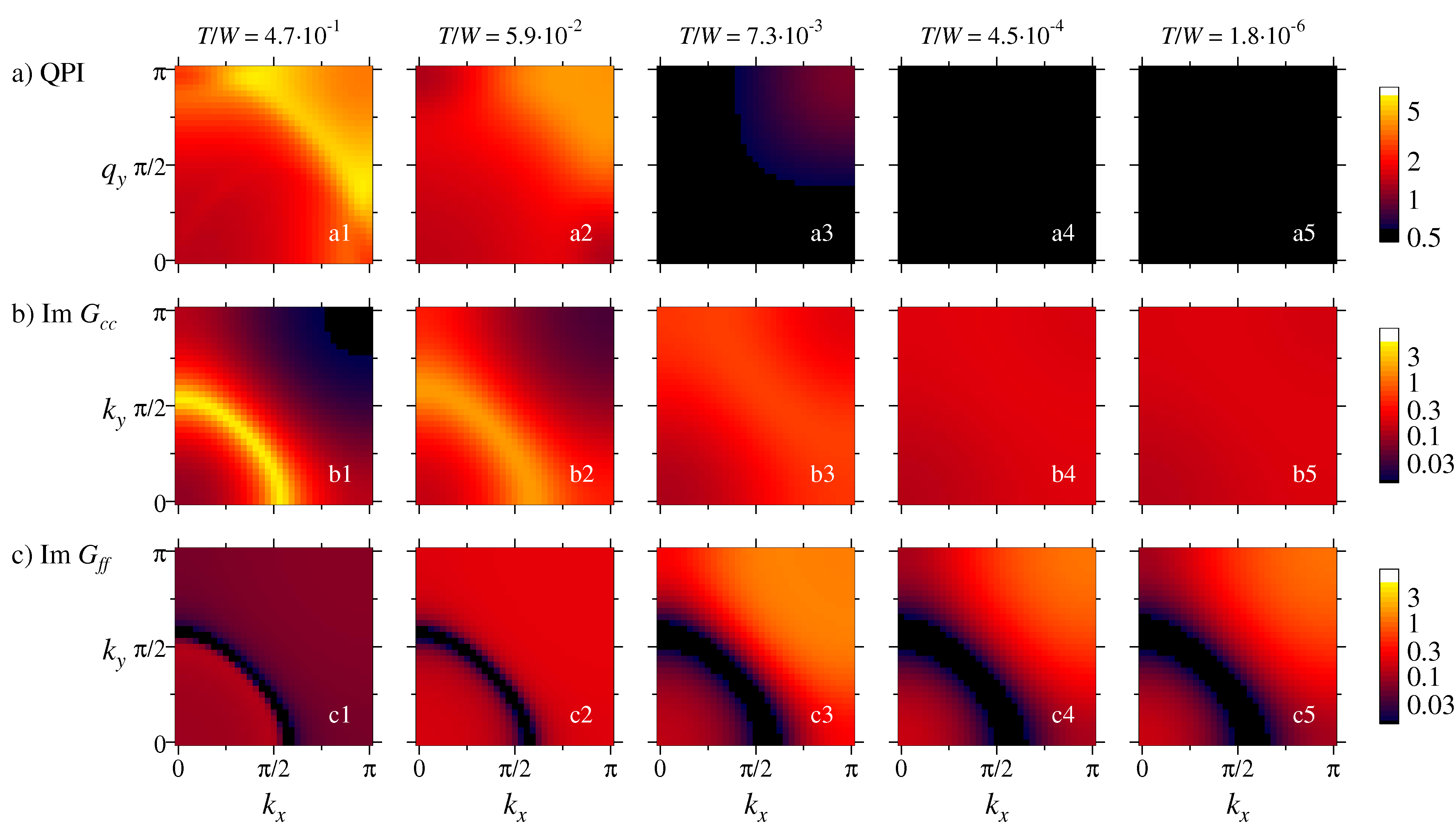}
\caption{
Same as in Fig.~\ref{qpi_fermi}, but for an energy of $\om=7.4 \times 10^{-3}$ inside the
hybridization gap.}
\label{qpi_gap}
\end{figure*}

An alternative way to represent QPI data and to discuss its relation to the
single-particle spectra is via constant-energy cuts. Such plots are displayed in
Figs.~\ref{qpi_fermi} and \ref{qpi_gap}, which show the QPI signal together with the $c$
and $f$ spectra for different temperature and energies $\om=0$ and $\om=4.7\times
10^{-3}$ (inside the hybridization gap), respectively.


At the Fermi level, Fig.~\ref{qpi_fermi}, the $c$-electron spectrum illustrates the
temperature evolution from the small Fermi surface at high temperature
(Fig.~\ref{qpi_fermi}b1) to the large Fermi surface at low temperature
(Fig.~\ref{qpi_fermi}b5). The $f$-electron spectrum displays a sharp Fermi surface only at
low $T$, where the intensity exceeds that in the $c$ spectrum, emphasizing that the
low-energy renormalized band has more $f$ than $c$ character.

The QPI signal has a strong intensity at a wave-vector $\bq$ essentially when the $c$
spectrum exhibits a sharp quasiparticle peak at $\bq/2$. In particular, the data at high
and low temperature, Figs.~\ref{qpi_fermi}a1 and \ref{qpi_fermi}a5, reflect the
corresponding Fermi surfaces. However, a unique extraction of the band structure from QPI
may be difficult: Figs.~\ref{qpi_fermi}a1 and \ref{qpi_fermi}a5 look rather similar, due
to the fact that the small Fermi surface has an electron volume of $n_c\approx 0.6$ while
the large Fermi surface has a {\em hole} volume $2-n_c-n_f\approx0.56$, i.e., both Fermi
surfaces yield similar wavevectors for elastic scattering. In actual STM experiments,
sub-atomic resolution allows one to obtain information beyond the first Brillouin zone,
such that those ambiguities can be partially resolved.
At intermediate temperatures $T\sim T_K$, the quasiparticle peak in the $c$ spectrum
dissolves, and consequently QPI response is very weak and diffuse,
Figs.~\ref{qpi_fermi}a2,a3.

%
%

It is instructive to compare the signal at the Fermi level, Fig.~\ref{qpi_fermi}, to that
at an energy inside the hybridization gap, Fig.~\ref{qpi_gap}. Again, at high
temperature, we have a sharp quasiparticle peak in the $c$ spectrum, corresponding to the
band forming a small Fermi surface, and, accordingly, an intense QPI response at the
wavevectors connecting portions of this iso-energy contour. Upon lowering the
temperature, both the quasiparticle peak and the QPI intensity decrease and essentially
disappear completely at the lowest temperature. This nicely reflects the absence of
well-defined quasiparticles inside the hybridization gap, i.e., all intensity inside this
pseudogap is incoherent.

%

\begin{figure*}[!t]
\centering
\includegraphics[width = 17.5cm]{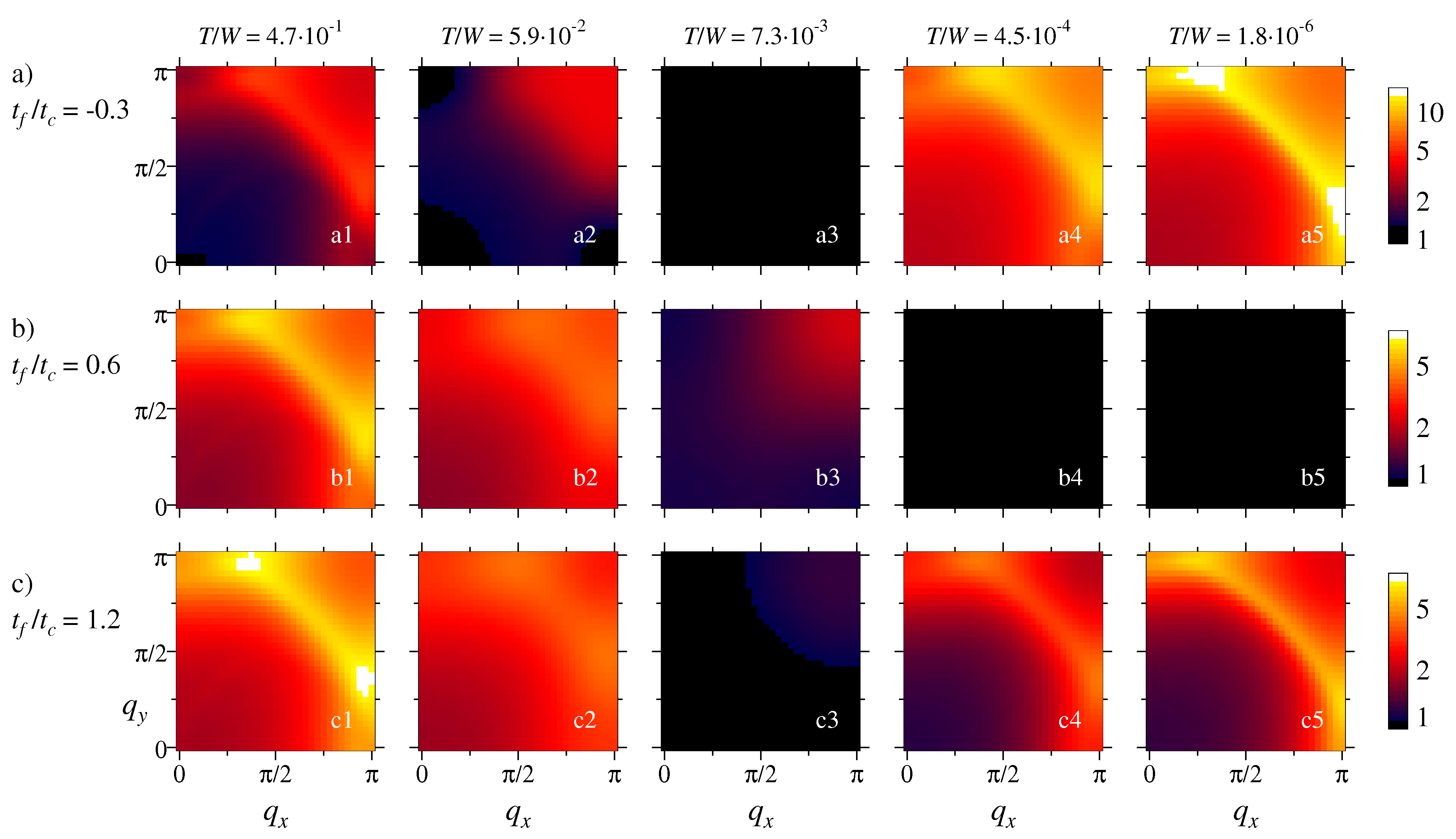}
\caption{
QPI signal at zero bias, $\om=0$, and different temperatures, now showing the variation
with the ratio of the tunneling amplitudes $t_f/t_c$.
a) $t_f/t_c = -0.3$,
b) $t_f/t_c = 0.6$,
c) $t_f/t_c = 1.2$.
The model parameters are the same as in Fig.~\ref{results:Sff_Scc}.
}
\label{qpi_tftc}
\end{figure*}


\subsubsection{Tunneling paths and QPI}

To complete our survey, we show the QPI signal at the Fermi level, $\om=0$, for a finite
ratio of the tunneling amplitudes $t_f/t_c$ in Fig.~\ref{qpi_tftc}. As discussed for the
differential conductance in the clean system (Sec.~\ref{subsec:dtc}), the destructive
(constructive) interference between the two tunneling channels for a positive (negative)
ratio $t_f/t_c$ is apparent through a reduction (enhancement) in the QPI signal at low
temperatures, while it is mainly unchanged at high temperature. The effect of the
destructive interference can be so strong that it essentially suppresses the QPI signal
at low energies and temperatures, as can be seen for $t_f/t_c=0.6$ in
Fig.~\ref{qpi_tftc}b.


\section{Summary}

We have analyzed the temperature-dependent electronic spectra in the periodic Anderson
model, a paradigmatic model for heavy-fermion formation, using dynamical mean field
theory (DMFT) with Wilson's numerical renormalization group (NRG). Particular attention
has been paid to the temperature evolution of the low-energy spectra and the crossover
from a `"small'' Fermi surface of conduction electrons to a "large'' Fermi surface of
composite heavy quasiparticles upon cooling. To make contact with STM experiments, we
have further studied the differential tunneling conductance, related to local spectral
properties of the model, and its modulations arising from impurity scattering processes
via quasiparticle interference (QPI). In particular, we go beyond the limitations of
previous analytical approaches, based on a slave boson mean-field theory, by fully
accounting for interaction-induced broadening of spectral features away from the Fermi
level and incoherent scattering processes at elevated temperature.

For the clean system, the local differential tunneling conductance is shown to display an
asymmetric peak, similar to the one observed experimentally, whose shape and intensity,
apart from material specific details, depend on the ratio of the tunneling amplitudes
$t_c$ and $t_f$ into conduction-electron and $f$-electron states, respectively.
(Previous theoretical studies found a hard gap near the Fermi level and recovered the
experimentally observed peak only by an \textit{ad hoc} addition of a phenomenological
quasiparticle broadening.) For positive ratio $t_f/t_c$, the effect of
destructive interferences between the two tunneling channels shows itself as a decrease
of the differential tunneling conductance intensity at low temperature, while it is
mainly inefficient at high temperature.

By studying at the momentum-dependent spectral functions at different temperatures, we
unveiled the {\em dynamical} reconstruction of the Fermi surface which happens mainly
within the optical gap. In this window of energies, spectral weight at high frequencies
is transferred from the bare conduction band to the edges of the renormalized bands when
the temperature is lowered down to zero. Islets of spectral weight at the Fermi level
persist but become smaller and smaller with decreasing temperatures until merging
completely with the incipient band structure in the $T\rightarrow 0$ limit. This
underlines the energy dependence of the crossover from a ``small'' to a ``large'' Fermi
surface: this crossover, characterized by dissolving quasiparticle peaks, happens at a
frequency-dependent temperature.

Quasiparticle interference induced by impurity scatterers offers an opportunity to follow
such a dynamical reconstruction, as one can at least partially reconstruct the band
structure from the dispersive high-intensity features of the QPI response. However, we have also shown
that this
may be seriously hindered by the destructive effect of the interference between the two
tunneling channels (here for positive ratio $t_f/t_c$), which may lead to an almost
complete extinction of QPI for certain energies.

Our detailed study spans many issues that are of relevance for existing and forthcoming
spectroscopic measurements on heavy-fermion materials. It would be particularly
interesting to study the energy-dependent crossover between small and large Fermi surfaces advocated
here. Layered heavy-fermion materials, e.g. of the so-called 115 family (CeCoIn$_5$ and
relatives) are most promising in this context, and corresponding experiments are
underway.\cite{yaz11}


\acknowledgments

The authors acknowledge fruitful discussions with F. Anders, J. C. S. Davis, D. K. Morr,
R. Peters, S. Wirth, and A. Yazdani. This research has been supported by the DFG through
GRK 1621 and FOR 960.



\begin{thebibliography}{}
\bibitem{te_note}
Strictly speaking, the tunneling of electrons from the tip to the system is a
non-equilibrium process. However, the amplitude of the current in STM experiments is
sufficiently small that the time between two tunneling events is much longer than the typical
electronic relaxation time. Then the thermal equilibrium assumption is appropriate.
\end{thebibliography}
\end{document}